\documentclass[journal]{IEEEtran}
\usepackage[table]{xcolor}
\usepackage{cite}
\usepackage{graphicx}
\usepackage{lipsum}
\usepackage{stfloats}
\usepackage{bm}
\usepackage{amsmath,amssymb}
\usepackage{amsthm}
\usepackage{pifont}
\usepackage{bbding}
\usepackage{booktabs}
\usepackage{ctable}
\usepackage{threeparttable}
\usepackage{footmisc}
\usepackage{framed} 
\usepackage{array}
\usepackage{multirow}
\usepackage{longtable}
\usepackage{rotating}
\usepackage{fancyhdr}
\usepackage{lastpage}
\usepackage{layout}
\usepackage{wrapfig}
\usepackage{tabularx}
\usepackage{tabu}
\usepackage{enumitem}
\usepackage{xcolor}
\usepackage{url}
\usepackage{dsfont}
\usepackage{arydshln}
\usepackage{algorithm}
\usepackage{algorithmic}
\usepackage{makecell}
\usepackage{enumerate}

\makeatletter
\newcommand{\removelatexerror}{\let\@latex@error\@gobble}
\makeatother



\newtheorem{myPropos}{\textbf{Proposition}}
\graphicspath{{Figures/}}
\DeclareGraphicsExtensions{.png, .eps}

\begin{document}
	\title{Joint Power and Spectrum Orchestration for D2D Semantic Communication Underlying Energy-Efficient Cellular Networks}
	\author{
	Le Xia,~\IEEEmembership{Member,~IEEE},
				 Yao Sun,~\IEEEmembership{Senior~Member,~IEEE},
				 Haijian Sun,~\IEEEmembership{Senior~Member,~IEEE},\\
				 Rose Qingyang Hu,~\IEEEmembership{Fellow,~IEEE},
				 Dusit Niyato,~\IEEEmembership{Fellow,~IEEE},
				 and Muhammad Ali Imran,~\IEEEmembership{Fellow,~IEEE}
	\thanks{
	
	
	Le Xia, Yao Sun (\textit{Corresponding author: Yao Sun}.), and Muhammad Ali Imran are with the James Watt School of Engineering, University of Glasgow, Glasgow G12 8QQ, UK (e-mail: xiale1995@outlook.com; \{Yao.Sun, Muhammad.Imran\}@glasgow.ac.uk).
	
	Haijian Sun is with the School of Electrical and Computer Engineering, University of Georgia, Athens, GA 30602, USA (e-mail: hsun@uga.edu).
	
	Rose Qingyang Hu is with the Bradley Department of Electrical and Computer Engineering, Virginia Tech, Blacksburg, VA 24061, USA (e-mail: rosehu@vt.edu).
	
	Dusit Niyato is with the College of Computing and Data Science, Nanyang Technological University, Singapore 639798 (e-mail: dniyato@ntu.edu.sg).
%
	
}
}	
	\maketitle
	\begin{abstract}
	Semantic communication (SemCom) has been recently deemed a promising next-generation wireless technique to enable efficient spectrum savings and information exchanges, thus naturally introducing a novel and practical network paradigm where cellular and device-to-device (D2D) SemCom approaches coexist.
	Nevertheless, the involved wireless resource management becomes complicated and challenging due to the unique semantic performance measurements and energy-consuming semantic coding mechanism.
	To this end, this paper jointly investigates power control and spectrum reuse problems for energy-efficient D2D SemCom cellular networks.
	Concretely, we first model the user preference-aware semantic triplet transmission and leverage a novel metric of semantic value to identify the semantic information importance conveyed in SemCom.
	Then, we define the additional power consumption from semantic encoding in conjunction with basic power amplifier dissipation to derive the overall system energy efficiency (semantic-value/Joule).
	Next, we formulate an energy efficiency maximization problem for joint power and spectrum allocation subject to several SemCom-related and practical constraints.
	Afterward, we propose an optimal resource management solution by employing the fractional-to-subtractive problem transformation and decomposition while developing a three-stage method with theoretical analysis of its optimality guarantee and computational complexity.
	 Numerical results demonstrate the adequate performance superiority of our proposed solution compared with different benchmarks.
	
	\end{abstract}
	

	\IEEEpeerreviewmaketitle
	
	\section{Introduction}
	\IEEEPARstart{R}{ecent} advances in semantic communication (SemCom) have shown great potential in enabling efficient information interaction and high resource utilization, promising to significantly relieve the scarcity of wireless resources in next-generation wireless cellular networks~\cite{zhang2022toward}.
	As a Shannon-beyond communication paradigm, SemCom concentrates upon accurately capturing the true meanings implied in source messages, rather than merely transmitting bits~\cite{10742565}.
	Particularly with the rapid advancement of artificial intelligence (AI), state-of-the-art and sophisticated deep learning (DL) models can be embedded into wireless devices to enable efficient, low-overhead transmission for a range of high-quality and large-capacity SemCom services, including both typical multimedia content (e.g., text~\cite{kadam2023knowledge}, image~\cite{erdemir2023generative}, and video~\cite{li2025goal}) and AI-generated content (AIGC)~\cite{du2024diffusion}.
	
	Notably, most of the related works have focused exclusively on device-to-device (D2D) SemCom, where the primary aim is to design DL-based semantic coding models and optimize semantic performance in D2D links.
	For example, Xie~\textit{et al.}~\cite{xie2021deep} devised a Transformer-based SemCom transceiver for reliable text transmission, which was then upgraded to be lightweight in~\cite{xie2020lite}.
	In~\cite{weng2021semantic}, Weng~\textit{et al.} employed an attention mechanism-enabled squeeze-and-excitation network for transmitting speech signals in D2D SemCom.
	Moreover, Xia~\textit{et al.}~\cite{xia2024generative} developed a generative AI-integrated end-to-end SemCom framework in a cloud-edge-mobile design for multimodal AIGC provisioning.
	While these lay the groundwork for implementing different D2D SemCom services, we notice that efficient semantic inference and recovery must rely on equivalent background knowledge and jointly trained coding models.
	Keeping this in mind, it is envisioned that D2D SemCom underlying cellular networks will be a very common and versatile architecture due to the following aspects.
	\begin{itemize}
		\item \textit{Spectrum Reuse and Capacity Gains:}
	By allowing D2D SemCom links to reuse cellular subcarriers, the network can carry both cellular user equipment (CUEs)' and D2D user equipments (DUEs)' semantic traffic on the same frequency bands to reduce spectrum fragmentation and increase overall throughput in semantics.
    For instance, if a CUE's subchannel is lightly loaded or can tolerate marginal interference, a DUE performing low-rate semantic exchanges (e.g., exchanging object labels in an AR application) can piggyback on that subchannel, thereby improving spectral efficiency without significantly degrading the cellular user's communication quality.
    \item \textit{Energy Efficiency and Latency Reduction:}
    It is known that the D2D links operate over much shorter distances than traditional uplink-downlink paths through the base station (BS)~\cite{jiang2019joint}, thereby generally incurring lower transmit power and propagation delay.
    Such features indicate that the each SemCom user only needs to consume fewer joules to transmit the same volume of semantic information compared to routing it through the BS, which can greatly enhance semantic-aware energy efficiency.
    \item \textit{Local Model Alignment and Robustness:}
    As aforementioned, effective SemCom normally requires transmitter and receiver to share background knowledge and align semantic coding models in time.
    In a pure cellular-only approach, each CUE has to upload raw sensor data to a central agent for joint model training, increasing both latency and backhaul load.
    By contrast, the D2D SemCom provides a quicker and more robust way for knowledge sharing and parameter training, also making the network more resilient to temporary cellular outages.
	\end{itemize}
	Accordingly, investigating D2D SemCom in cellular networks can significantly improve the efficiency and robustness of SemCom itself, while retaining cellular network control capabilities to better meet the future needs of diverse, high-speed, low-latency, and economical wireless networks.
	
	As a matter of fact, there have been some noteworthy technical works addressing a variety of challenges in pure cellular SemCom networks.
	Yan~\textit{et al.}~\cite{yan2022resource} investigated the semantic spectral efficiency optimization-based channel assignment within the base station.
	Zhang~\textit{et al.}~\cite{zhang2024semantic} proposed a multi-agent reinforcement learning-based, semantic-aware power and channel allocation scheme for cellular vehicle-to-everything platooning.
	The multi-modal SemCom-based intelligent resource allocation was then considered by Hu~\textit{et al.}~\cite{hu2025resource} for multi-unmanned aerial vehicles relay collaboration.
	To further blend semantic and traditional bit communications, Xia \textit{et al.}~\cite{xia2024wireless} developed a best mode selection strategy for multiple cellular users with joint optimization of user association and bandwidth allocation for hybrid semantic/bit communication networks.
	Likewise, Evgenidis \textit{et al.}~\cite{10477313} proposed a single-cell multi-carrier hybrid Semantic-Shannon system aiming to minimize the transmission delay.
	In addition, the resource management problems in conventional cellular-D2D networks have also been fully explored.
	In~\cite{sun2018uplink}, Sun \textit{et al.} adopted an interference limited area based D2D management scheme for mitigating interference on cellular links.
	Pawar \textit{et al.} maximized the network throughput for both cellular and D2D users in a joint uplink-downlink manner in~\cite{pawar2021joint}.
	Nevertheless, to the best of our knowledge, none of the existing work has ever focused on the~\textit{D2D SemCom Network} (D2D-SCN) from an energy-efficient networking perspective, in which cellular SemCom users and D2D SemCom user pairs coexist.
	
	In full view of the energy-efficient D2D-SCN, our main task lies in seeking the optimal strategy for two closely relevant and coupled wireless resource management issues of power control and spectrum reuse in it.
	Unfortunately, solutions to conventional communication scenarios cannot be directly applied due to the new focus of semantic delivery and the new requirements of semantic coding in SemCom.
	To be more specific, SemCom entails more circuit power consumption for semantic coding, while SemCom users expect only high semantic fidelity in line with their individual semantic service preferences.
	This is because existing semantic models typically involve sophisticated neural network operations that require additional computation for AI-driven semantic inference~\cite{zhang2022toward}, increasing circuit power versus conventional bit-level coding.
	Especially when considering energy-efficient large-scale D2D-SCNs, how to maximize the overall system energy efficiency (\textit{semantics-per-Joule}) by trading off energy consumption and acquired semantic-level performance at multiple cellular and D2D SemCom users should be rather complicated and challenging.
	These considerations give rise to three fundamental challenges in D2D-SCNs:
	\begin{itemize}
		\item \textit{Challenge 1: How to measure the semantic-level performance for each wireless SemCom link?} Differing from traditional bit-oriented communication, the semantic-level performance needs to be carefully characterized in SemCom due to its sole focus on meaning delivery. Especially noting that users may have personal preferences and background knowledge for varying SemCom services, even the same messages transmitted from different SemCom users could represent distinct semantic importance, which raises the first nontrivial point unique to SemCom scenarios compared to the conventional cellular-D2D scenarios.
		\item \textit{Challenge 2: How to adequately consider the semantic coding process when defining the new energy efficiency model in SemCom?} Owing to unique semantic encoding/decoding requirements, additional power for information extraction and inference is consumed in SemCom-enabled transceivers. This shifts the energy efficiency analysis from bits-per-Joule to semantics-per-Joule, which is also not considered in classical D2D or cellular energy models, and thus requiring a new definition of energy efficiency for D2D-SCN.
		\item \textit{Challenge 3: How to achieve the best energy-efficient D2D-SCN through wireless resource optimization?} Clearly, power allocation and spectrum reuse are two key issues in resource utilization that can significantly affect the achievable semantic performance for each SemCom user. If aiming at maximizing overall energy efficiency, besides practical energy models' features and constraints, each user's semantic service demand along with its individual preference must also be taken into account in resource optimization, thereby posing the third difficulty.
	\end{itemize}
	
	It should be noted that the challenges listed above are specific to D2D-SCNs rather than generic to other network paradigms, since none of the existing works on non-semantic cellular-D2D coexistence or semantic-only D2D setups jointly address user-preference-aware semantic fidelity, semantic coding cost, and semantics-pertain energy efficiency across both CUEs and DUEs.
	In response to these challenges, in this paper, we propose an optimal joint power allocation and spectrum reuse strategy for energy efficiency maximization in D2D-SCNs taking into account the unique SemCom characteristics.
	Both theoretical analysis and numerical results showcase the performance superiority of the proposed solution in terms of energy efficiency, semantic performance, and total power consumption compared with two different benchmarks.
	In a nutshell, our main contributions are summarized as follows:
	\begin{itemize}
		\item We first introduce a fine‐grained ``semantic triplet'' as the unit of transmission in SemCom, tightly coupled to each CUE's and DUE's individual semantic service preferences.
		 We then leverage a novel metric called semantic value as the Zipf-weighted count of semantic triplets delivered per second, quantifying semantic information importance.
		 This addresses the \textit{Challenge 1}.
		\item We carefully define the power consumption occurred during the semantic encoding process in conjunction with basic power amplifier dissipation to derive the overall energy efficiency of D2D-SCN. This semantics-centric cost model contrasts sharply with traditional models that consider only bits-per-Joule and ignore semantic inference costs. Afterward, we formulate a joint power and spectrum orchestration NP-hard problem to maximize the semantic value-based energy efficiency subject to several SemCom-related and practical system constraints. The contribution directly addresses \textit{Challenge 2}.
		\item We develop an efficient resource management solution to address the optimization problem, and its optimality is theoretically proved by two propositions. Specifically, a fractional-to-subtractive transformation approach is employed to decompose the complex primal problem into multiple tractable subproblems. Then, a three-stage method is devised to solve each subproblem with polynomial-time computational complexity. In each iteration of the solution, the first and second stages are to obtain the optimal power allocation policy and the third stage is to finalize the optimal spectrum reusing pattern. In this way, \textit{Challenge 3} is finally well tackled.
	\end{itemize}

	\begin{table}[t]
		\centering
		\caption{List of Main Notations}
		\label{ListNotation}
		\setlength{\tabcolsep}{3pt}
		\renewcommand\arraystretch{1.3}
		\begin{tabular}{|m{1.1cm}<{\raggedright}|m{7.1cm}<{\raggedright}|}\hline
			\textbf{Notation} & \textbf{Description} \\ \hline
			$\mathcal{M}$ & The set of CUEs, each CUE $i \in \mathcal{M}=\{1,2,\cdots,M\}$\\ \hline
			$\mathcal{N}$ & The set of DUEs, each DUE $j \in \mathcal{N}=\{1,2,\cdots,N\}$\\ \hline
			$\mathcal{K}$ & The set of SemCom services, each SemCom service $k \in \mathcal{K}=\{1,2,\cdots,K\}$\\ \hline
			$P_{i}^{C}$ & Transmit power of CUE $i$, no greater than $P_{\mathit{max}}^{C}$ \\ \hline
			$P_{j}^{D}$ & Transmit power of DUE $j$, no greater than $P_{\mathit{max}}^{D}$ \\ \hline
			$\alpha_{i,j}$ & Binary spectrum reuse indicator between CUE $i$ and DUE $j$ \\ \hline
			$L$ & Average number of bits required to encode a semantic triplet \\ \hline
			$W$ & Bandwidth allocated to the orthogonal uplink of each CUE\\ \hline
			$V_{i}^{C}$ & Semantic value at CUE $i$, no less than $V_{min}^{C}$\\ \hline
			$V_{j}^{D}$ & Semantic value at DUE $j$, no less than $V_{min}^{D}$\\ \hline
			$P^{\mathit{enc}}$ & Circuit power required to encode a semantic triplet\\ \hline
			$\xi$ & Power amplifier inefficiency coefficient\\ \hline
			$V^{\mathit{total}}$ & Semantic value transmitted by all SemCom users\\ \hline
			$E^{\mathit{total}}$ & Power consumption at all SemCom users\\ \hline
			$\eta_{\mathit{EE}}$ & Energy efficiency of the D2D-SCN\\ \hline
			$F(\eta_{\mathit{EE}})$ & Supremum of the subtractive-form function of $\eta_{\mathit{EE}}$\\ \hline
			$\lambda_{i,j}$ & Sum term in the subtractive-form function of $\eta_{\mathit{EE}}$, pertain to a single spectrum-reuse pair of CUE $i$-DUE $j$\\ \hline
			$\psi$ & Feasible region of the subproblem after problem decomposition, pertain to every single spectrum-reuse pair\\ \hline
			$\check{\lambda}_{i}$ & Sum term in the subtractive-form function of $\eta_{\mathit{EE}}$, pertain to a single CUE $i$ without spectrum reuse\\ \hline
			$Q$ & The maximum number of iterations for updating $\eta_{\mathit{EE}}$\\ \hline
		\end{tabular}
	\end{table}
	The remainder of this paper is organized as follows.
	Section II first introduces the system model of D2D-SCN and formulates the associated energy efficiency maximization problem.
	Then, we illustrate the proposed optimal power allocation and spectrum reusing strategy in Section III.
	Numerical results are demonstrated and discussed in Section IV, followed by the conclusions in Section V.
	In addition, important notations used in the paper are listed in Table~\ref{ListNotation} for better readability.

    \section{System Model and Problem Formulation}
	In this section, the considered D2D-SCN scenario is first elaborated along with the semantic performance measurement and semantic coding-related energy efficiency model. Then, the corresponding resource optimization problem is presented. 
	 
	\subsection{D2D-SCN Scenario}
\begin{figure}[t]
		\centering
		\includegraphics[width=0.48\textwidth]{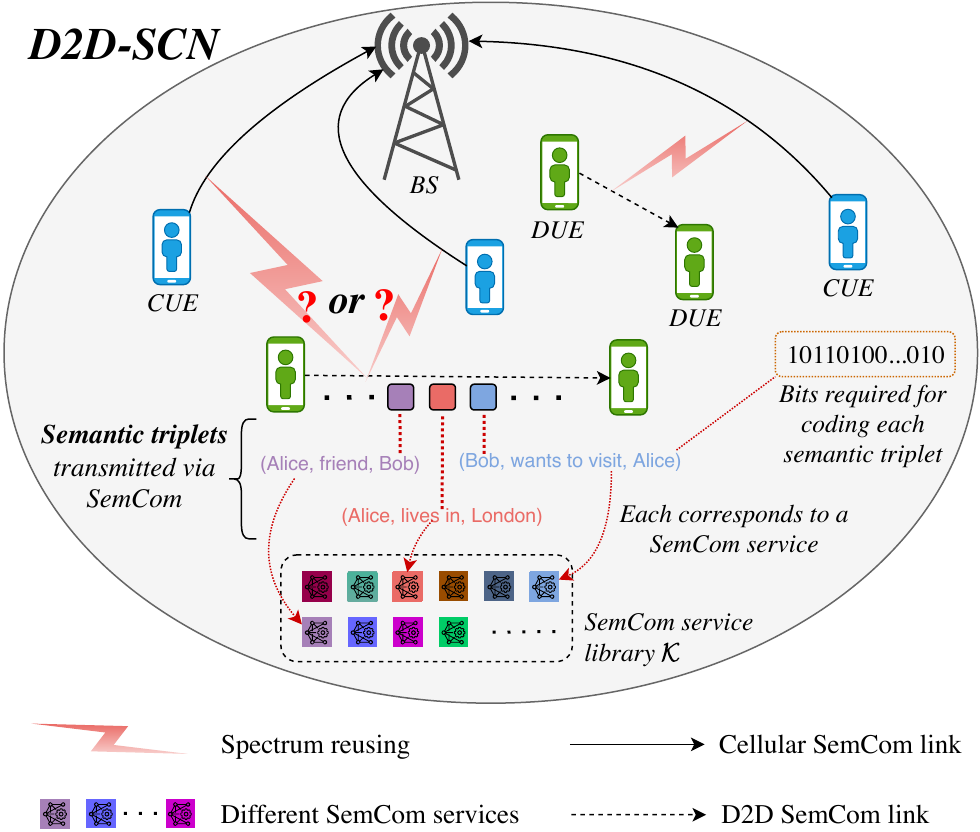} 
		\caption{The D2D-SCN with multiple SemCom-enabled CUEs and DUEs.}
		\label{Scenario}
    \end{figure}

	Consider a single-cell D2D-SCN scenario as shown in Fig.~\ref{Scenario}, where $M$ CUEs and $N$ $(N \leqslant M)$\footnote{This modeling is consistent with the classical one-to-one spectrum reuse assumption, which has been widely used in related works~\cite{pawar2021joint,guo2019resource}. For the case of $N>M$, it is served by a different way of system modeling, which is beyond the scope of this work and can be included in the extended study.} pairs of DUEs are capable of performing $K$ different wireless SemCom services\footnote{Here, different SemCom services can be deemed different semantic delivery tasks, each based on a certain modality (e.g., text or image) associated with a specific knowledge domain (e.g., sports or music)~\cite{10742565,xia2023xurllc}.} based on a SemCom service library $\mathcal{K}=\{1,2,\cdots,K\}$.
	Each CUE $i \in \mathcal{M}=\{1,2,\cdots,M\}$ is assumed to be pre-allocated an orthogonal uplink subchannel with equal channel bandwidth $W$ to execute SemCom with its remote receiver.
	In addition, each DUE $j \in \mathcal{N}=\{1,2,\cdots,N\}$ is allowed to reuse the subchannel of only one CUE for SemCom service provisioning, and to preserve generality, the subchannel of each CUE can be reused by at most one DUE.
	Let $\alpha_{i,j}\in \{0,1\}$ denote the spectrum reuse indicator, where $\alpha_{i,j}=1$ represents that DUE $j$ reuses the subchannel of CUE $i$, and $\alpha_{i,j}=0$ otherwise.
	Furthermore, assuming that each CUE and each DUE have their respective maximum transmit power, denoted by $P_{\mathit{max}}^{C}$ and $P_{\mathit{max}}^{D}$.
	
	\subsection{Channel Model and Semantic Performance Measurement}
	In the data dissemination model, the channel power gain between CUE $i$ and the BS, the gain between DUE $j$ and the BS, the gain between the transmitter and the receiver at DUE $j$, and the gain between each CUE $i$ and each DUE $j$ are first denoted as $G_{i, B}$, $G_{j, B}$, $G_{j}^{D}$, and $G_{i,j}$, respectively.
	Notably, the path loss and fading-related factors are treated as constant and have been implied in the above channel power gains.
	Moreover, due to the system modeling in the previous subsection, the only cross-link interference seen by CUE $i$ only comes from the single DUE with $\alpha_{i,j} = 1$.
	Likewise, DUE $j$ only experiences interference from its paired CUE $i$ when $\alpha_{i,j} = 1$.
	With these, if denoting the transmit power of each CUE $i$ and DUE $j$ as $P_{i}^{C}$ and $P_{j}^{D}$, respectively, the achievable instantaneous bit rate at the uplink of CUE $i$ is
	\begin{equation}
		\label{cueSINR}
		\begin{aligned}
			r_{i}^{C} = W\log_{2}\left(1+\frac{P_{i}^{C}G_{i, B}}{\delta^{2}+\sum_{j\in \mathcal{N}}\alpha_{i,j}P_{j}^{D}G_{j, B}}\right),
		\end{aligned}
	\end{equation}
	and the achievable instantaneous bit rate at DUE $j$ is
	\begin{equation}
		\label{dueSINR}
		\begin{aligned}
			r_{j}^{D} = W\log_{2}\left(1+\frac{P_{j}^{D}G_{j}^{D}}{\delta^{2}+\sum_{i\in \mathcal{M}}\alpha_{i,j}P_{i}^{C}G_{i, j}}\right),
		\end{aligned}
	\end{equation}
	where $\delta^{2}$ is the noise power.

	As for the SemCom model, a concept of \textit{semantic triplet} is introduced to represent the interpretable relationship between two specific semantic entities implied in source information~\cite{yang2019transms,chen2020review,gao2024importance}, and its typical expression is~\textit{(Entity-A, Relationship, Entity-B)}, for instance, (Alice, friend, Bob), as depicted in Fig.~\ref{Scenario}.
	In this work, we assume that different SemCom services randomly arrive at CUEs and DUEs~\cite{10261329}, and the source information corresponding to arbitrary SemCom service needs to be fed into the semantic encoder to extract multiple semantic triplets.\footnote{This is justified as core meanings can be extracted in the form of semantic triplets (via DL methods or knowledge graphs, etc.) for multimedia SemCom service provisioning (e.g., text~\cite{gao2024importance}, image~\cite{shi2021semantic}, and video~\cite{xia2023wiservr}).}
	Each semantic triplet is then encoded into an average of $L$ bits on average by each user's channel encoder and transmitted over the wireless channel~\cite{gao2024importance}.
	Such a bit-level modeling allows us to quantify semantic-level link capacity expressions in a straightforward manner without introducing additional, unnecessary system parameters.
	As such, the total number of semantic triplets that can be transmitted by CUE $i$ and DUE $j$ per second are given by $\lfloor r_{i}^{C}/L\rfloor$ and $\lfloor r_{j}^{D}/L\rfloor$, respectively, where $\lfloor \cdot \rfloor$ is the floor function.
	
	Meanwhile, it is noticed that different CUEs and DUEs may have different personal preferences for these $K$ SemCom services, resulting in varying proportions of the number of semantic triplets based on different SemCom services during transmission.
	Here, assuming that the information contained in each semantic triplet is associated with a particular type of SemCom service, and intuitively, the more popular a type of SemCom service is, the higher the probability that the corresponding semantic triplets will be encoded.
	Without the loss of generality, we consider that the SemCom service popularity at each CUE and DUE follows the Zipf distribution~\cite{piantadosi2014zipf},\footnote{Other known skewed probability distributions can also be adopted, however, Zipf offers a more realistic and mathematically convenient baseline.} and all CUEs and DUEs are assumed to have identical semantic encoding capabilities for any source information.
	The rationale behind the assumption of Zipf distribution is because it is a well-established model for capturing skewed user preferences in real-world systems, particularly in wireless content delivery, web access, and video streaming.
    In these settings, a small subset of services or content types tends to dominate user demand, while the majority are accessed infrequently--a distribution pattern also expected in SemCom scenarios, where users repeatedly request SemCom services (e.g., specific text, image, or AIGC tasks) aligned with their personal interests or tasks.
    This approach has also been applied in our prior SemCom work~\cite{xia2023xurllc}.
	In this way, if these SemCom services are ranked by decreasing popularity (for example, the most popular SemCom service $k \in \mathcal{K}$ at CUE $i$ has the rank $u_{i,k}^{C} = 1$, the next one has $u_{i,k}^{C} = 2$, etc.), then the expected number of semantic triplets transmitted for SemCom service $k$ per second is calculated by $\lfloor r_{i}^{C}/L\rfloor \cdot (u_{i,k}^{C})^{-\beta_{i}^{C}}/\sum_{e\in \mathcal{K}}e^{-\beta_{i}^{C}}$, where $\beta_{i}^{C}$ ($\beta_{i}^{C}\geqslant 0$) is the skewness of CUE $i$'s Zipf distribution.\footnote{The SemCom service popularity ranking of each user can be analyzed and estimated based on its historical messaging records~\cite{hassine2015popularity,xia2023xurllc,li2018service}.}
	Likewise, the expected number of semantic triplets transmitted for SemCom service $k$ at DUE $j$ per second is $\lfloor r_{j}^{D}/L\rfloor \cdot (u_{j,k}^{D})^{-\beta_{j}^{D}}/\sum_{e\in \mathcal{K}}e^{-\beta_{j}^{D}}$, where $u_{j,k}^{D}$ and $\beta_{j}^{D}$ are DUE $j$'s popularity ranking for SemCom service $k$ and the skewness of its Zipf distribution, respectively.
	Notably, user-specific preferences simply weight the encoding probability of semantic triplets for each SemCom service, which is independent of the aforementioned random service arrivals.
	Although service arrivals are ``random in time'', the choice of which service is used for semantic triplet generation is biased toward higher‐ranked services in exactly the way described by the Zipf distribution.
	
	Naturally, transmitting the higher-ranked semantic triplets contributes more valuable semantic information for each SemCom receiver.
	Inspired by this, we employ a performance metric called~\textit{semantic value} proposed in~\cite{gao2024importance} to measure the semantic information importance of semantic triplets with different rankings.\footnote{Note that other similar semantic-level metrics can also be applied here without changing the remaining modeling and solution.}
	According to~\cite{gao2024importance}, the semantic value of one semantic triplet is precisely defined by $(u_{i,k}^{C})^{-\beta_{i}^{C}}$, where $u_{i,k}^{C}$ and $\beta_{i}^{C}$ have been described before in the context of the Zipf distribution.
	Such a definition is justified by well-established observations that semantic importance correlates with frequency in natural language~\cite{piantadosi2014zipf}.
	In other words, more ``central'' or ``informative'' concepts naturally occur more frequently, thus producing a similar heavy-tailed Zipf's profile.
	This is completely consistent with human intuition, i.e., the more critical semantics during communication receive substantially higher weight.
	As such, the Zipf distribution can be leveraged again for semantic value measurement.
	
	Specifically, CUE $i$'s each semantic triplet encoded based on SemCom service $k$ with the rank $u_{i,k}^{C}$ has the semantic value of $(u_{i,k}^{C})^{-\beta_{i}^{C}}$, and that of DUE $j$ has the semantic value of $(u_{j,k}^{D})^{-\beta_{j}^{D}}$.
	Combined with the number of their transmitted semantic triplets as obtained before, the semantic value transmitted by CUE $i$ per second should be:
	\begin{equation}
		V_{i}^{C} =\left\lfloor \frac{r_{i}^{C}}{L}\right\rfloor \sum_{u_{i,k}^{C}\in \mathcal{K}}\frac{(u_{i,k}^{C})^{-2\beta_{i}^{C}}}{\sum_{e\in \mathcal{K}}e^{-\beta_{i}^{C}}} \triangleq \theta_{i}^{C}\left\lfloor \frac{r_{i}^{C}}{L}\right\rfloor,\label{cueSV}
	\end{equation}
	and the semantic value transmitted by DUE $j$ per second is
	\begin{equation}
		V_{j}^{D} =\left\lfloor \frac{r_{j}^{D}}{L}\right\rfloor \sum_{u_{j,k}^{D}\in \mathcal{K}}\frac{(u_{j,k}^{D})^{-2\beta_{j}^{D}}}{\sum_{e\in \mathcal{K}}e^{-\beta_{j}^{D}}}\triangleq \theta_{j}^{D}\left\lfloor \frac{r_{j}^{D}}{L}\right\rfloor,\label{dueSV}
	\end{equation}
	where the two parameters $\theta_{i}^{C}$ and $\theta_{j}^{D}$ are defined for brevity.
	Clearly, the overall semantic value transmitted by all CUEs and DUEs in the D2D-SCN per second is given by $V^{\mathit{total}} = \sum_{i \in \mathcal{M}}V_{i}^{C}+\sum_{j \in \mathcal{N}}V_{j}^{D}$.
	It is further required that $V_{i}^{C}\geqslant V_{min}^{C}$ and $V_{j}^{D}\geqslant V_{min}^{D}$, where $V_{min}^{C}$ and $V_{min}^{D}$ are the unified minimum thresholds for each CUE and each DUE, respectively.
	
	\subsection{Energy Efficiency Model of SemCom Systems}
	In this work, we focus on the overall power consumption including the contributions of the unique semantic encoding circuit module and the transmit power amplifier at each SemCom user.
	Different from conventional models focusing on the circuit power consumption on bit processing~\cite{guo2017energy} or antenna controlling~\cite{ng2012energy}, the power consumption in the SemCom-enabled user devices is primly considered to be related to the computing for each semantic triplet during the semantic encoding process.\footnote{Note that the circuit power consumption of receiver-side semantic decoding is not considered as the receiver deals with the actual semantic content effectively received and should be distinguished from the resource scheduling policy at the transmitter side. Furthermore, the initial focus of this work is on energy efficiency at the uplink side of D2D-SCN, as depicted in Fig. 1, and thus only the transmit power is the main concern.}
	This is justified because the relationship between two information entities in each semantic triplet necessarily requires a certain computing power to be accurately identified, reasoned and interpreted, and such semantic encoding tasks are sometimes accomplished by sophisticated DL algorithms-integrated circuit module~\cite{liew2022economics}.
	In this way, the semantic value, encoding cost, and energy efficiency are together coupled with power and spectrum orchestration since the number of semantic triplets that can be transmitted fundamentally depends on the designed resource allocation strategies.
	
	Now, suppose that all CUEs and DUEs have fixed circuit power consumption, denoted by $P^{\mathit{enc}}$, to encode each semantic triplet, and assume that all semantic triplets generated by the semantic encoder per second are fully transmitted out.
	Hence, the total circuit power consumption (in \textit{Joule/s}) for semantic encoding in the D2D-SCN can be estimated by
	\begin{equation}
		\label{powerSC}
			E^{S} = P^{\mathit{enc}}\left(\sum_{i \in \mathcal{M}}\left\lfloor \frac{r_{i}^{C}}{L}\right\rfloor+\sum_{j \in \mathcal{N}}\left\lfloor \frac{r_{j}^{D}}{L}\right\rfloor\right).
	\end{equation}
	
	In addition, most of related works also consider the power dissipation at power amplifiers of transmitters~\cite{auer2011much,jiang2019joint,guo2017energy}, which has been widely recognized as one of the energy loss sources in present wireless networks.
	We define the power amplifier inefficiency coefficient as $\xi$ ($\xi \geqslant 1$), which is a constant associated with the transmit power of each CUE and DUE.
	As such, the total power consumption for semantic triplet transmission in the D2D-SCN is obtained from
	\begin{equation}
		\label{powerAmpli}
		\begin{aligned}
			E^{T} = \xi\left(\sum_{i \in \mathcal{M}}P_{i}^{C}+\sum_{j \in \mathcal{N}}P_{j}^{D}\right).
		\end{aligned}
	\end{equation}
	
	Accordingly, the total power consumption at all CUEs and DUEs is $E^{\mathit{total}} = E^{S}+E^{T}$.
	Since the semantic value performance becomes the sole focus of SemCom, we define the energy efficiency of D2D-SCN as the overall semantic value successfully transferred to the SemCom-enabled receiver per Joule of energy consumed on average, given by
	\begin{equation}
		\begin{aligned}
			\eta_{\mathit{EE}} &= \frac{V^{\mathit{total}}}{E^{\mathit{total}}}=\frac{\sum_{i \in \mathcal{M}}V_{i}^{C}+\sum_{j \in \mathcal{N}}V_{j}^{D}}{E^{S}+E^{T}}.
		\end{aligned}
	\end{equation}
 	
	\subsection{Problem Formulation}
	For ease of illustration, we first define three variable sets $\bm{P^{C}}=\left\{P_{i}^{C}\mid i \in \mathcal{M}\right\}$, $\bm{P^{D}}=\left\{P_{j}^{D}\mid j \in \mathcal{N}\right\}$, and $\bm{\alpha}=\left\{\alpha_{i,j}\mid i \in \mathcal{M}, j \in \mathcal{N}\right\}$ that consist of all possible indicators pertinent to power allocation and spectrum reusing, respectively.
	Without loss of generality, the objective is to maximize $\eta_{\mathit{EE}}$ of D2D-SCN by jointly optimizing $(\bm{P^{C}},\bm{P^{D}}, \bm{\alpha})$, and subject to SemCom-related requirements alongside several system constraints.
	The problem is formulated as follows:
	\begin{align}
	\mathbf{P0}:\ \max_{\bm{P^{C}},\bm{P^{D}}, \bm{\alpha}} \quad & \eta_{\mathit{EE}}~\label{P0}\\
	{\rm s.t.} \quad & V_{i}^{C}\geqslant V_{min}^{C},\ \forall i\in \mathcal{M},\tag{\ref{P0}a}\\
	& V_{j}^{D}\geqslant V_{min}^{D},\ \forall j\in \mathcal{N},\tag{\ref{P0}b}\\
	& 0\leqslant P_{i}^{C}\leqslant P_{max}^{C},\ \forall i\in\mathcal{M},\tag{\ref{P0}c}\\
	& 0\leqslant P_{j}^{D}\leqslant P_{max}^{D},\ \forall j\in\mathcal{N},\tag{\ref{P0}d}\\
	& \sum_{j\in\mathcal{N}}\alpha_{i,j}\leqslant 1,\ \forall i\in\mathcal{M},\tag{\ref{P0}e}\\
	& \sum_{i\in\mathcal{M}}\alpha_{i,j}=1,\ \forall j \in\mathcal{N},\tag{\ref{P0}f}\\
	& \alpha_{i,j}\in \{0,1\}, \ \forall \left( i,j\right) \in\mathcal{M}\times \mathcal{N}\tag{\ref{P0}g}.
	\end{align}
	Constraints (\ref{P0}a) and (\ref{P0}b) guarantee the minimum semantic value achieved at each CUE and DUE, respectively.
	Similarly, constraints (\ref{P0}c) and (\ref{P0}d) limit the maximum transmit power for each CUE and DUE, respectively.
	Constraint (\ref{P0}e) represents that the subchannel of each CUE can be reused by at most one DUE, while some remaining CUEs (due to $N \leqslant M$) may not need to share its spectrum with DUEs.
	Constraint (\ref{P0}f) requires that each DUE can reuse only one subchannel of an existing CUE at one time, as aforementioned.
	Finally, constraint (\ref{P0}g) characterizes the binary property of $\bm{\alpha}$.

	Carefully examining $\mathbf{P0}$, it can be observed that the optimization is quite challenging to be solved straightforwardly due to several intractable mathematical obstacles.
	First, $\mathbf{P0}$ is an NP-hard optimization problem, as demonstrated below.
	We can consider a simplified case of $\mathbf{P0}$, where all $(\bm{P^{C}}, \bm{P^{D}})$-related constraints are already satisfied and $V_{min}^{C}=V_{min}^{D}=0$ is set.
	As such, constraints (\ref{P0}a)-(\ref{P0}d) can all be removed and $\mathbf{P0}$ degenerates into a typical $0$-$1$ integer programming problem that is known to be NP-hard~\cite{lenstra1983integer}, not even to mention that the objective $\eta_{\mathit{EE}}$ remains distinctly nonlinear.
	Hence, $\mathbf{P0}$ is also NP-hard.
	In addition, the expression of $\eta_{\mathit{EE}}$ is quite complicated alongside the constraints (\ref{P0}a) and (\ref{P0}b), which are nonconvex and thus generally requires a high-complexity solution procedure.
	Therefore, we propose an efficient power allocation and spectrum reusing strategy in the next section to reach the optimality of $\mathbf{P0}$.
	
	\section{Proposed Resource Allocation for D2D-SCNs}
\begin{figure*}[ht]
		\centering
		\includegraphics[width=0.8\textwidth]{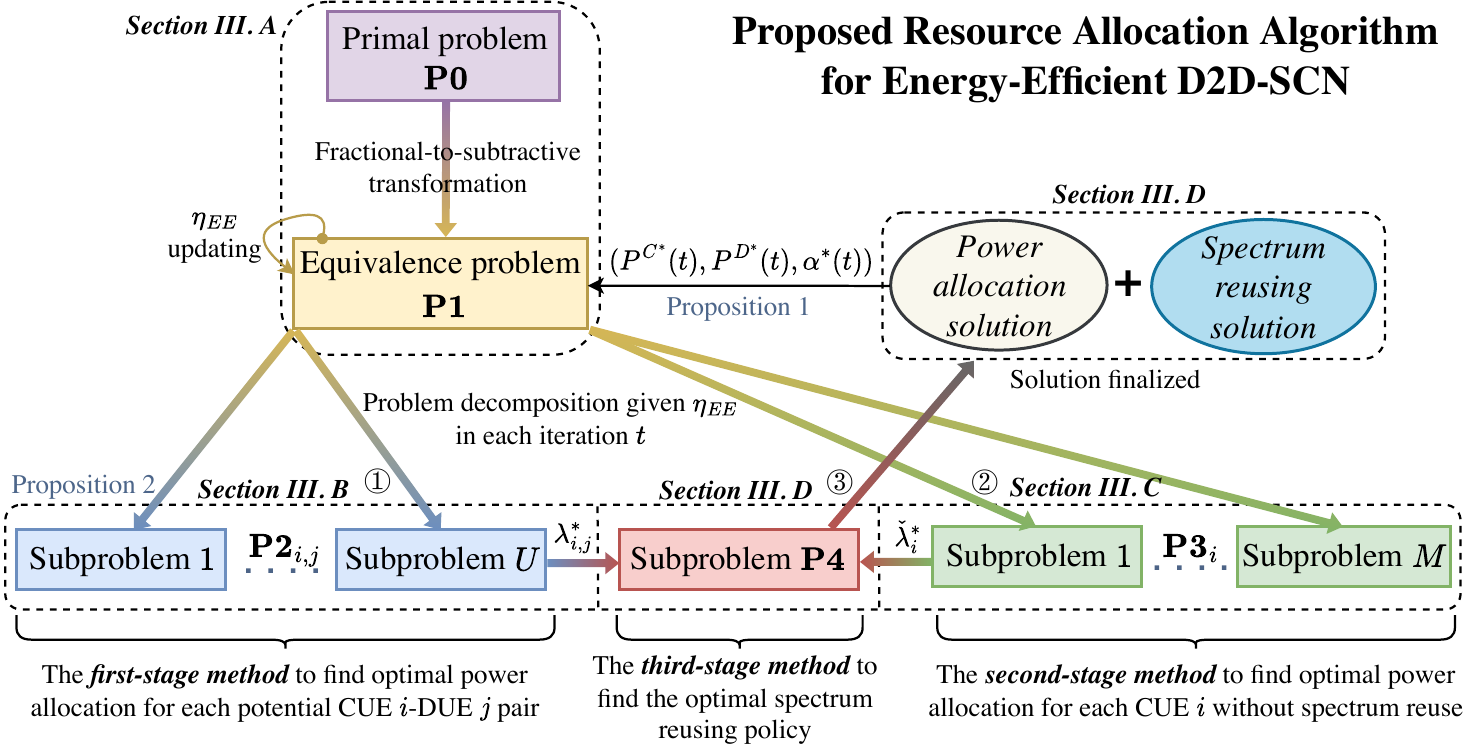} 
		\caption{Illustration of the proposed resource allocation algorithm.}
		\label{Solution}
    \end{figure*}

	In this section, we illustrate how to design our resource allocation strategy to cope with the energy efficiency optimization problem in the D2D-SCN.
	As depicted in Fig.~\ref{Solution}, the primal problem $\mathbf{P0}$ is first transformed without losing optimality, from its original fractional form into an equivalent subtractive form (referring to $\mathbf{P1}$ in Subsection III-A) by employing the Dinkelbach's method~\cite{dinkelbach1967nonlinear}.
	Then, $\mathbf{P1}$ is decomposed into multiple subproblems that will be solved in an iterative fashion, and in each iteration, we specially devise a three-stage method.
	In the first and the second stage, we construct $U=M\times N$ subproblems (referring to $\mathbf{P2}_{i, j},\forall \left( i,j\right) \in\mathcal{M}\times \mathcal{N}$, in Subsection III-B) and $M$ subproblems (referring to $\mathbf{P3}_{i},\forall i \in \mathcal{M}$, in Subsection III-C), respectively.
	Among them, each $\mathbf{P2}_{i, j}$ corresponds to the power allocation problem for a potential spectrum reusing pair of CUE $i$ and DUE $j$, while each $\mathbf{P3}_{i}$ corresponds to that for each CUE $i$ without spectrum reusing.
	After solving all these subproblems, the spectrum reusing policy with respect to (w.r.t.) $\bm{\alpha}$ is finalized (referring to $\mathbf{P4}$ in Subsection III-D).
	In the end, we present the workflow of our proposed solution along with its complexity analysis in Subsection III-E.
	
	\subsection{Fractional-to-Subtractive Problem Transformation}
	Since the objective $\eta_{\mathit{EE}}$ in the primal problem $\mathbf{P0}$ is complex, we employ Dinkelbach's method~\cite{dinkelbach1967nonlinear} to transform it into the subtractive form $F\left(\eta_{\mathit{EE}}\right)$, rendering the problem decomposable and tractable.
	Given $\eta_{\mathit{EE}}$ as a known value, the transformation is established by the following proposition.
	\begin{myPropos}
		\label{PropC2_1}
		$\mathbf{P0}$ must have the same optimal solution as
		\begin{align}
			\mathbf{P1}:\ F\left(\eta_{\mathit{EE}}\right)=&\max_{\bm{P^{C}},\bm{P^{D}}, \bm{\alpha}} \quad  V^{\mathit{total}}-\eta_{\mathit{EE}}\cdot
			E^{\mathit{total}}~\label{P1}\\
			{\rm s.t.} \quad & \text{(\ref{P0}a)}-\text{(\ref{P0}g)},\tag{\ref{P1}a}
		\end{align}
	if and only if $F\left(\eta_{\mathit{EE}}\right)=0$.
	\end{myPropos}
	\begin{IEEEproof}
		Please see Appendix A.
	\end{IEEEproof}
	
	From Proposition~\ref{PropC2_1}, it is seen that when $F\left(\eta_{\mathit{EE}}\right)$, i.e., the optimal value of $\mathbf{P1}$, is equal to $0$, the optimal solution to $\mathbf{P1}$ also constitutes an optimal solution to $\mathbf{P0}$.
	Hence, our optimization goal is to solve $\mathbf{P1}$ given any $\eta_{\mathit{EE}}$ while requiring the iterative update for $\eta_{\mathit{EE}}$ such that $F\left(\eta_{\mathit{EE}}\right)$ eventually approaches $0$, and thus reaching the optimality of $\mathbf{P0}$.
	Specifically, we begin by assigning an arbitrary non-negative initial value to $\eta_{\mathit{EE}}$ for $\mathbf{P1}$, which needs to be updated in each iteration according to the following rule:
	\begin{equation}
	\label{update}
		\eta_{\mathit{EE}}(t+1)=\frac{V^{\mathit{total}}\left(\bm{P^{C^{*}}}(t),\bm{P^{D^{*}}}(t), \bm{\alpha^{*}}(t)\right)}{E^{\mathit{total}}\left(\bm{P^{C^{*}}}(t),\bm{P^{D^{*}}}(t), \bm{\alpha^{*}}(t)\right)},
	\end{equation}
	where $\left(\bm{P^{C^{*}}}(t),\bm{P^{D^{*}}}(t), \bm{\alpha^{*}}(t)\right)$ is the optimal solution to $\mathbf{P1}$ in iteration $t$.
	$V^{\mathit{total}}\left(\cdot\right)$ and $E^{\mathit{total}}\left(\cdot\right)$ represent the functional forms of $V^{\mathit{total}}$ and $E^{\mathit{total}}$ w.r.t. the variables $\left(\bm{P^{C}},\bm{P^{D}}, \bm{\alpha}\right)$, respectively.
	Note that such iterative update should be stopped when either reaching the maximum number of iterations (denoted by $Q$) or satisfying a condition of $F\left(\eta_{\mathit{EE}}(t)\right)<\epsilon$, where $\epsilon$ is a preset small positive value~\cite{guo2017energy}.
	Most importantly, it has been proved that if $Q$ is large enough, the convergence of $\eta_{\mathit{EE}}$ can be guaranteed such that the optimality condition in Proposition~\ref{PropC2_1} is satisfied, i.e., $F\left(\eta_{\mathit{EE}}(t)\right)=0$, and the details of proof can refer to~\cite{ng2012energy} and~\cite{dinkelbach1967nonlinear}.
	
	Given any $\eta_{\mathit{EE}}$ in each iteration, we now concentrate upon how to reach the optimality of $\mathbf{P1}$.
	However, solving such a problem is still tricky due to the mixed integer variables in its highly complex objective function~(\ref{P1}).
	To this end, $\mathbf{P1}$ will be first decomposed into multiple subproblems, and then we specially propose a three-stage method to separately obtain the optimal power allocation scheme $\left(\bm{P^{C}},\bm{P^{D}}\right)$ and the optimal spectrum reusing policy $\bm{\alpha}$ with polynomial-time complexity.

	\subsection{Power Allocation for A Spectrum-Sharing CUE-DUE Pair}
	In the first stage, the power allocation scheme is optimized for a specific pair of CUE $i$ ($\forall i \in \mathcal{M}$) and DUE $j$ ($\forall j \in \mathcal{N}$).
	As such, we construct $U=M\times N$ subproblems, each of which is denoted as $\mathbf{P2}_{i,j}$ and the objective is to maximize energy efficiency for the single spectrum reusing pair.
	It is worth pointing out that combining the optimal single-pair solutions to these $\mathbf{P2}_{i,j}$ cannot directly achieve the optimal power allocation strategy for $\mathbf{P1}$, but these solutions will be used to construct the subsequent spectrum reuse subproblem so as to help finalize the resource allocation strategy for $\mathbf{P1}$.
	Accordingly, when DUE $j$ reuses the subchannel of CUE $i$ (i.e., $\alpha_{i,j}=1$), given any $\eta_{\mathit{EE}}$, $\mathbf{P2}_{i,j}$ becomes
	\begin{align}
			\mathbf{P2}_{i,j}:\ \max_{P_{i}^{C},P_{j}^{D}} \quad &  \lambda_{i,j}~\label{P2}\\
			{\rm s.t.} \quad & V_{i}^{C}\geqslant V_{min}^{C},\tag{\ref{P2}a}\\
			&V_{j}^{D}\geqslant V_{min}^{D},\tag{\ref{P2}b}\\
			&0\leqslant P_{i}^{C}\leqslant P_{max}^{C},\tag{\ref{P2}c}\\
			&0\leqslant P_{j}^{D}\leqslant P_{max}^{D}.\tag{\ref{P2}d}
	\end{align}
	In particular, $\lambda_{i,j}$ is defined as the sum of all the terms related to the fixed pair of CUE $i$ and DUE $j$ in $F\left(\eta_{\mathit{EE}}\right)$, given by~(\ref{rewrittedobjective}) at the bottom of this page, in which $\sigma_{i}^{C}$ and $\sigma_{j}^{D}$ are defined for brevity.
	Here, we decompose the complex expression of $\lambda_{i,j}$ and combine like terms about the variables to facilitate subsequent optimization analysis.
	\begin{figure*}[hb]
		\centering
		\hrulefill
		\begin{equation}
			\label{rewrittedobjective}
			\begin{aligned}
				\lambda_{i,j}&=\theta_{i}^{C}\left\lfloor \frac{\overline{r_{i}^{C}}}{L}\right\rfloor+\theta_{j}^{D}\left\lfloor \frac{\overline{r_{j}^{D}}}{L}\right\rfloor -\eta_{\mathit{EE}} \left[P^{\mathit{enc}}\left(\left\lfloor \frac{\overline{r_{i}^{C}}}{L}\right\rfloor+\left\lfloor \frac{\overline{r_{j}^{D}}}{L}\right\rfloor\right) +\xi\left(P_{i}^{C} + P_{j}^{D}\right)\right]\\
			&=\left(\theta_{i}^{C}-\eta_{\mathit{EE}}P^{\mathit{enc}}\right)\left\lfloor \frac{\overline{r_{i}^{C}}}{L}\right\rfloor+\left(\theta_{j}^{D}-\eta_{\mathit{EE}}P^{\mathit{enc}}\right)\left\lfloor \frac{\overline{r_{j}^{D}}}{L}\right\rfloor -\eta_{\mathit{EE}}\xi\left(P_{i}^{C}+ P_{j}^{D}\right)\\
			&\triangleq \sigma_{i}^{C}\left\lfloor \frac{\overline{r_{i}^{C}}}{L}\right\rfloor+\sigma_{j}^{D}\left\lfloor \frac{\overline{r_{j}^{D}}}{L}\right\rfloor-\eta_{\mathit{EE}}\xi\left(P_{i}^{C}+ P_{j}^{D}\right).
			\end{aligned}
		\end{equation}
	\end{figure*}
 	Notably, $\overline{r_{i}^{C}}$ and $\overline{r_{j}^{D}}$ newly introduced in~(\ref{rewrittedobjective}) are calculated by substituting $\alpha_{i,j}=1$ into~\eqref{cueSINR} and~\eqref{dueSINR}, respectively, which can be found as
	\begin{equation}
	\label{ricbar}
		\overline{r_{i}^{C}}=W\log_{2}\left(1+\frac{P_{i}^{C}G_{i, B}}{\delta^{2}+P_{j}^{D}G_{j, B}}\right)
	\end{equation}
	and
	\begin{equation}
	\label{rjdbar}
		\overline{r_{j}^{D}}=W\log_{2}\left(1+\frac{P_{j}^{D}G_{j}^{D}}{\delta^{2}+P_{i}^{C}G_{i, j}}\right).
	\end{equation}
	
	
	
	\begin{figure*}[ht]
		\centering
		\includegraphics[width=0.95\textwidth]{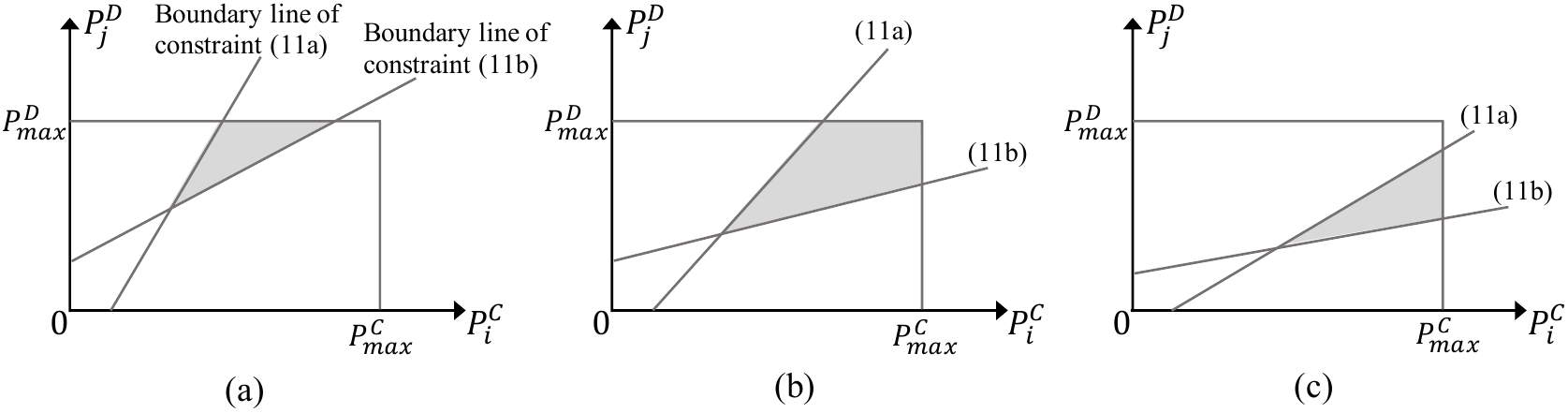} 
		\caption{Three possible cases of the feasible power allocation region $\psi$ for each pair of CUE $i$ and DUE $j$ w.r.t. $\mathbf{P2}_{i,j}$.}
		\label{feasiregion}
    \end{figure*}
	Note that $\theta_{i}^{C}$ and $\theta_{j}^{D}$ are two positive parameters as in~\eqref{cueSV} and~\eqref{dueSV}, and thus constraints~(\ref{P2}a) and~(\ref{P2}b) in $\mathbf{P2}_{i,j}$ can be smoothly transformed to $\overline{r_{i}^{C}}\geqslant L\left\lceil V_{min}^{C}/\theta_{i}^{C}\right\rceil$ and $\overline{r_{j}^{D}}\geqslant L\left\lceil V_{min}^{D}/\theta_{j}^{D}\right\rceil$, respectively.
	By further considering the boundary case of either of the two constraints, according to~\eqref{ricbar} and~\eqref{rjdbar}, $P_{j}^{D}$ can clearly be expressed as a linear function of $P_{i}^{C}$.
	Fig.~\ref{feasiregion} depicts three possible cases for the closed feasible region $\psi$ w.r.t. $\left(P_{i}^{C},P_{j}^{D}\right)$, which can help us to roughly determine the shape of the solution space.
	Now, if we first generate an arbitrary feasible solution in $\psi$, denoted as $\left(\widetilde{P_{i}^{C}},\widetilde{P_{j}^{D}}\right)$, then considering a special case of substituting $\left(\widetilde{P_{i}^{C}},\widetilde{P_{j}^{D}}\right)$ into the first term $\sigma_{i}^{C}\left\lfloor \overline{r_{i}^{C}}/L\right\rfloor$ in~\eqref{rewrittedobjective}, such that
	\begin{equation}
	\label{initialpoint}
		\sigma_{i}^{C}\left\lfloor \frac{W}{L}\log_{2}\left(1+\frac{\widetilde{P_{i}^{C}}G_{i, B}}{\delta^{2}+\widetilde{P_{j}^{D}}G_{j, B}}\right)\right\rfloor\triangleq \lambda_{0},
	\end{equation}
	where $\lambda_{0}$ is denoted as the corresponding solved value.
	Based on~\eqref{initialpoint}, there must be a line segment w.r.t. $\left(P_{i}^{C},P_{j}^{D}\right)\in \psi$ passing through the point $\left(\widetilde{P_{i}^{C}},\widetilde{P_{j}^{D}}\right)$, which can be expressed as a linear equation of $P_{i}^{C}$ w.r.t. $P_{j}^{D}$, i.e.,
	\begin{equation}
	\label{straightline}
	\begin{aligned}
		P_{i}^{C}&=\frac{G_{j, B}\left(2^{\frac{L}{W}\left\lceil\lambda_{0}/\sigma_{i}^{C}\right\rceil}-1\right)}{G_{i, B}}P_{j}^{D}\!+\!\frac{\delta^{2}\left( 2^{\frac{L}{W}\left\lceil\lambda_{0}/\sigma_{i}^{C}\right\rceil}-1\right)}{G_{i, B}}\\
		&\triangleq k_0P_{j}^{D}+b_0,
	\end{aligned}
	\end{equation}
	where $k_0$ and $b_0$ are denoted as constants.	
	Clearly, any point $\left(P_{i}^{C},P_{j}^{D}\right)$ on~\eqref{straightline} results in the term $\sigma_{i}^{C}\left\lfloor \overline{r_{i}^{C}}/L\right\rfloor$ being computed as $\lambda_{0}$.
	Keeping this in mind, we then concentrate upon the remaining terms of $\lambda_{i,j}$.
	By employing \eqref{straightline}, we can substitute $P_{i}^{C}$ in~\eqref{rewrittedobjective} by $P_{j}^{D}$, such that
	\begin{equation}
	\label{remainterms}
		\begin{aligned}
			\lambda_{i,j}&= \sigma_{j}^{D}\left\lfloor \overline{r_{j}^{D}}/L\right\rfloor-\eta_{\mathit{EE}}\xi\left(P_{i}^{C}+ P_{j}^{D}\right)+\lambda_{0}\\
			&=\sigma_{j}^{D}\!\left\lfloor \! \frac{W}{L}\log_{2}\!\left(\!1+\frac{G_{j}^{D}}{\left[\left(\delta^{2}+G_{i, j}b_0\right)/P_{j}^{D}\right]+G_{i, j}k_0}\!\right)\!\right\rfloor \\
			&\ \quad \quad-\eta_{\mathit{EE}}\xi\left[\left(k_0+1\right)P_{j}^{D}+b_0\right]+\lambda_{0}\\
			&\triangleq \sigma_{j}^{D}\left\lfloor \phi_1\left(P_{j}^{D}\right) \right\rfloor+\phi_2\left(P_{j}^{D}\right)+\lambda_{0}\\
			&\triangleq \widetilde{\lambda_{\mathit{1}}}\left(P_{j}^{D}\right)+\lambda_{0}.
		\end{aligned}
	\end{equation}
	Here, $\phi_1\left(P_{j}^{D}\right)$ is the logarithmic function of $P_{j}^{D}$, $\phi_2\left(P_{j}^{D}\right)$ is the linear function of $P_{j}^{D}$, and $\widetilde{\lambda_{\mathit{1}}}\left(P_{j}^{D}\right)$ is the combinatorial function of $\phi_1\left(P_{j}^{D}\right)$ and $\phi_2\left(P_{j}^{D}\right)$.
	
	Note that $\phi_1\left(P_{j}^{D}\right)$ is increasing w.r.t. $P_{j}^{D}$, and hence $\sigma_{j}^{D}\left\lfloor \phi_1\left(P_{j}^{D}\right) \right\rfloor$ clearly becomes a monotone staircase function of $P_{j}^{D}$, and its monotonicity depends on the positivity or negativity of $\sigma_{j}^{D}$.
	Combined with the linear property of $\phi_2\left(P_{j}^{D}\right)$, it is easily found that the optimal $P_{j}^{D}$ leading to the maximal $\lambda_{i,j}$ must be exactly one endpoint of a particular staircase of $\sigma_{j}^{D}\left\lfloor \phi_1\left(P_{j}^{D}\right) \right\rfloor$.
	For instance, if $\sigma_{j}^{D}\geqslant 0$ and $-\eta_{\mathit{EE}}\xi\left(k_0+1\right) \geqslant 0$, the optimal $P_{j}^{D}$ must be the rightmost endpoint of the rightmost staircase of $\sigma_{j}^{D}\left\lfloor \phi_1\left(P_{j}^{D}\right) \right\rfloor$.
	Moreover, considering the closed feasible region $\psi$, the number of staircases of $\sigma_{j}^{D}\left\lfloor \phi_1\left(P_{j}^{D}\right) \right\rfloor$ must be limited, thereby a brute-force search~\cite{grothendieck2013non} can be applied to determine the optimal $P_{j}^{D}$.
	Afterward, the optimal $P_{i}^{C}$ should be obtained based on~\eqref{straightline}.
	In summary, over the set of points on the line segment obtained by fixing the first term in \eqref{rewrittedobjective}, we can always find the optimal $\left(P_{i}^{C},P_{j}^{D}\right)$ to maximize $\lambda_{i,j}$.
	
	Next, if we substitute the same solution $\left(\widetilde{P_{i}^{C}},\widetilde{P_{j}^{D}}\right)$ into the second term $\sigma_{j}^{D}\left\lfloor \overline{r_{j}^{D}}/L\right\rfloor$ in~\eqref{rewrittedobjective}, another line segment can be determined similar to the rationale behind~\eqref{initialpoint}.
	Different with \eqref{straightline}, this line segment is expressed as a linear equation of $P_{j}^{D}$ w.r.t. $P_{i}^{C}$.
	Given that, substituting $P_{j}^{D}$ in~\eqref{rewrittedobjective} by $P_{i}^{C}$ can yield an expression form of $\lambda_{i,j}$ w.r.t. $P_{i}^{C}$, denoted by $\lambda_{i,j}\triangleq \widetilde{\lambda_{\mathit{2}}}\left(P_{i}^{C}\right)+\lambda'_{0}$.
	Likewise, $\widetilde{\lambda_{\mathit{2}}}\left(P_{i}^{C}\right)$ is the combinatorial function of one staircase function and one linear function w.r.t. $P_{i}^{C}$.
	By again leveraging the brute-force search, we can also find the optimal $\left(P_{i}^{C},P_{j}^{D}\right)$ over the set of points on the line segment obtained by fixing the second term in \eqref{rewrittedobjective}.	
	
	In view of the above, the following proposition shows how $\widetilde{\lambda_{\mathit{1}}}\left(P_{j}^{D}\right)$ and $\widetilde{\lambda_{\mathit{2}}}\left(P_{i}^{C}\right)$ influence the optimality of $\mathbf{P2}_{i,j}$.
	\begin{myPropos}
	\label{PropC2_2}
		In $\psi$, given any known feasible solution $\left(\widetilde{P_{i}^{C}},\widetilde{P_{j}^{D}}\right)$ of $\mathbf{P2}_{i,j}$, let $\left(\overleftarrow{P_{i}^{C}},\overleftarrow{P_{j}^{D}}\right)$ be the optimal point on the line segment that maximizes $\widetilde{\lambda_{\mathit{1}}}\left(P_{j}^{D}\right)$ as in~\eqref{remainterms}, where the line segment is determined by substituting $\left(\widetilde{P_{i}^{C}},\widetilde{P_{j}^{D}}\right)$ into $\sigma_{i}^{C}\left\lfloor \overline{r_{i}^{C}}/L\right\rfloor$ as in~\eqref{straightline}.
		Likewise, let $\left(\overrightarrow{P_{i}^{C}},\overrightarrow{P_{j}^{D}}\right)$ be the optimal point on another line segment that maximizes $\widetilde{\lambda_{\mathit{2}}}\left(P_{i}^{C}\right)$, where the line segment is determined by substituting $\left(\widetilde{P_{i}^{C}},\widetilde{P_{j}^{D}}\right)$ into $\sigma_{j}^{D}\left\lfloor \overline{r_{j}^{D}}/L\right\rfloor$.
		Then, the optimal solution to $\mathbf{P2}_{i,j}$ must satisfy
		\begin{equation}
			\left(P_{i}^{C^*}\!,P_{j}^{D^*}\!\right)\!\in\! \left\{\!\left(\overleftarrow{P_{i}^{C}},\overleftarrow{P_{j}^{D}}\right)\!\mid\! \left(\overleftarrow{P_{i}^{C}},\overleftarrow{P_{j}^{D}}\right)\!=\!\left(\overrightarrow{P_{i}^{C}}\!,\overrightarrow{P_{j}^{D}}\right)\!\in \!\psi\right\}.\label{ggwa}
		\end{equation}
	\end{myPropos}
	\begin{IEEEproof}
	Please see Appendix B.
	\end{IEEEproof}
	From Proposition~\ref{PropC2_2}, it is seen that $\mathbf{P2}_{i,j}$'s optimal solution $\left(P_{i}^{C^*},P_{j}^{D^*}\right)$ must belong to the set of points on the right side of \eqref{ggwa}.
	In other words, the solution point should exactly be the coincide point that makes $\widetilde{\lambda_{\mathit{1}}}\left(P_{j}^{D}\right)$ and $\widetilde{\lambda_{\mathit{2}}}\left(P_{i}^{C}\right)$ reach their respective maxima at the same time.
	In line with this, we specially devise a heuristic search algorithm to efficiently determine the power allocation strategy for each possible CUE-DUE pair.
	In detail, our power allocation solution is realized by the following five phases:
	\begin{itemize}
		\item \textit{(1) Initial Feasible Solution Generation:} First, we need to generate an initial feasible solution $\left(\widetilde{P_{i}^{C}},\widetilde{P_{j}^{D}}\right)$ as the search starting point.
		Straightforwardly, it can be set as arbitrary corner point of $\psi$.
		For instance, by combining (\ref{P2}a) and (\ref{P2}b) with \eqref{ricbar} and \eqref{rjdbar}, if $\psi$ appears to the case of Fig.~\ref{feasiregion}(a), $\left(\widetilde{P_{i}^{C}},\widetilde{P_{j}^{D}}\right)$ can set to be $\left(\left(2^{\frac{L}{W}\left\lceil V_{min}^{C}/\theta_{i}^{C}\right\rceil}-1\right)\left(\delta^{2}+P_{max}^{D}G_{j, B}\right)/G_{i, B}, P_{max}^{D}\right)$, be $\left(P_{max}^{C},\left(2^{\frac{L}{W}\left\lceil V_{min}^{D}/\theta_{j}^{D}\right\rceil}-1\right)\left(\delta^{2}+P_{max}^{C}G_{i,j}\right)/G_{j}^{D}\right)$ if in Fig.~\ref{feasiregion}(c), and be $\left(P_{max}^{C},P_{max}^{D}\right)$ if in Fig.~\ref{feasiregion}(b).
		\item \textit{(2) Optimal Line Point Search for Maximum $\widetilde{\lambda_{\mathit{1}}}\left(P_{j}^{D}\right)$:} By executing the procedures in \eqref{initialpoint}-\eqref{remainterms}, the close-form expression of $\widetilde{\lambda_{\mathit{1}}}\left(P_{j}^{D}\right)$ is obtained.
		Note that the domain of $P_{j}^{D}$ should be constrained by the line segment as in \eqref{straightline}.
		On this basis, by directly substituting the minimum and maximum $P_{j}^{D}$ into $\sigma_{j}^{D}\left\lfloor \phi_1\left(P_{j}^{D}\right) \right\rfloor$, respectively, the values and the two endpoints of all staircases can be obtained, to which the brute-force search is applied to determine $\left(\overleftarrow{P_{i}^{C}},\overleftarrow{P_{j}^{D}}\right)$ and it is not difficult to solve.
%
		\item \textit{(3) Optimal Line Point Search for Maximum $\widetilde{\lambda_{\mathit{2}}}\left(P_{i}^{C}\right)$:} We substitute the obtained $\left(\overleftarrow{P_{i}^{C}},\overleftarrow{P_{j}^{D}}\right)$ into the second term $\sigma_{j}^{D}\left\lfloor \overline{r_{j}^{D}}/L\right\rfloor$ in~\eqref{rewrittedobjective} to obtain the close-form expressions of its corresponding line segment and $\widetilde{\lambda_{\mathit{2}}}\left(P_{i}^{C}\right)$.
		By substituting the minimum and maximum $P_{i}^{C}$ and again leveraging the brute-force search, $\left(\overrightarrow{P_{i}^{C}},\overrightarrow{P_{j}^{D}}\right)$ is obtained.
		\item \textit{(4) Searching Termination Check:} If $\left(\overleftarrow{P_{i}^{C}},\overleftarrow{P_{j}^{D}}\right)=\left(\overrightarrow{P_{i}^{C}},\overrightarrow{P_{j}^{D}}\right)$, i.e., the optimal point of the previous line segment is also optimal for the current searching line segment, this round of search is terminated.
		Otherwise, the obtained $\left(\overrightarrow{P_{i}^{C}},\overrightarrow{P_{j}^{D}}\right)$ is substituted into the first term $\sigma_{i}^{C}\left\lfloor \overline{r_{i}^{C}}/L\right\rfloor$ in~\eqref{rewrittedobjective}, and then keep repeating \textit{Phases (2)} and \textit{(3)} until $\left(\overleftarrow{P_{i}^{C}},\overleftarrow{P_{j}^{D}}\right)=\left(\overrightarrow{P_{i}^{C}},\overrightarrow{P_{j}^{D}}\right)$ is satisfied.
		According to Proposition~\ref{PropC2_2}, the coincide point may still fall into a local optimum of $\mathbf{P2}_{i,j}$, and hence we add it into a list, denoted by $\mathcal{I}=\left\{\!\left(\overleftarrow{P_{i}^{C}},\overleftarrow{P_{j}^{D}}\right)\!\mid\! \left(\overleftarrow{P_{i}^{C}},\overleftarrow{P_{j}^{D}}\right)\!=\!\left(\overrightarrow{P_{i}^{C}}\!,\overrightarrow{P_{j}^{D}}\right)\!\in \!\psi\right\}$, for record.
		\item \textit{(5) Multiple Rounds of Searches:} Setting \textit{Phases (1)-(4)} as a single round of search, multiple rounds of search are repeated until reaching a preset maximum round restriction, and in different rounds, the initial feasible solution $\left(\widetilde{P_{i}^{C}},\widetilde{P_{j}^{D}}\right)$ should always be different to traverse all local optima as much as possible.
		Due to the nonconvex and non-linear nature of the objective $\lambda_{i,j}$ and feasible region $\psi$, only the sub-optimal power allocation strategy $\left(P_{i}^{C^*},P_{j}^{D^*}\right)$ is finalized by comparing all the solutions recorded in $\mathcal{I}$.
		Notably, our approach trades off computational complexity for tractability, which is a very common and acceptable strategy for NP-hard resource allocation problems in wireless networks~\cite{10032275,guo2017energy}.
	\end{itemize}
	
	It is worth noting that the above iterative round search must eventually contribute to convergence for the following two reasons.
	The first is because the associated feasible region $\psi$ is finite as in Fig.~\ref{feasiregion}, and technically there exists at least one point that leads to the maximum $\lambda_{i,j}$.
	The second reason lies in a fact that every search in \textit{Phases (2)} or \textit{(3)} is oriented towards maximizing $\lambda_{i,j}$, as explained earlier.
	To be more concrete, each repetition of \textit{Phases (2)} or \textit{(3)} should always yield a larger $\lambda_{i,j}$ by computing \eqref{rewrittedobjective} if $\left(\overleftarrow{P_{i}^{C}},\overleftarrow{P_{j}^{D}}\right)\neq\left(\overrightarrow{P_{i}^{C}},\overrightarrow{P_{j}^{D}}\right)$, however, the value of $\lambda_{i,j}$ is obviously not infinite, and thus such search cannot be endless.
	Therefore, the proposed iterative search procedure is bound to converge to one point that meets the condition in \eqref{ggwa} within a finite number of steps.
	
	\subsection{Power Allocation for each CUE Without Spectrum Sharing}
	Since the preset number of DUEs does not exceed that of CUEs (i.e., $N \leqslant M$), there must be a part of CUEs' sub-channels not reused by any DUE.
	Accordingly, it is also necessary to determine the optimal power allocation solution for each CUE $i$ without spectrum sharing, and thus our second-stage problem $\mathbf{P3}_{i}$ ($\forall i \in \mathcal{M}$) becomes
	\begin{align}
			\mathbf{P3}_i:\ \max_{P_{i}^{C}} \quad &  \check{\lambda}_{i}~\label{P3}\\
			{\rm s.t.} \quad & \theta_{i}^{C}\left\lfloor \check{r_{i}}^{C}/L\right\rfloor\geqslant V_{min}^{C},\tag{\ref{P3}a}\\
			&\text{(\ref{P2}c)},\tag{\ref{P3}b}
	\end{align}
	where $\check{r_{i}}^{C}=W\log_{2}\left(1+\frac{P_{i}^{C}G_{i, B}}{\delta^{2}}\right)$, representing the achievable bit rate at CUE $i$ when $\alpha_{i,j}=0$, and $\check{\lambda}_{i} = \sigma_{i}^{C}\left\lfloor \check{r_{i}}^{C}/L\right\rfloor-\eta_{\mathit{EE}}\xi P_{i}^{C}$, indicating all the terms related to only CUE $i$ in~\eqref{P1}.
	
	Combined constraint (\ref{P3}a) with (\ref{P3}b), the domain of feasible $P_{i}^{C}$ is confined as $\left[\delta^{2}\left(2^{\frac{L}{W}\left\lceil V_{min}^{C}/\theta_{i}^{C}\right\rceil}-1\right)/G_{i, B}\right] \leqslant P_{i}^{C} \leqslant P_{max}^{C}$.
	With this, it is further observed that the first term $\sigma_{i}^{C}\left\lfloor \check{r_{i}}^{C}/L\right\rfloor$ in $\check{\lambda}_{i}$ is a monotone staircase function of $P_{i}^{C}$, while its second term $-\eta_{\mathit{EE}}\xi P_{i}^{C}$ is a linear function of $P_{i}^{C}$.
	Recap our previous analysis to~\eqref{remainterms}, the optimal $P_{i}^{C}$ to each $\mathbf{P3}_i$ can be obtained by again using the brute-force search with acceptable computation complexity, due to the simplified problem structure and the limited scale of the search space.
	
	
	\subsection{Optimal Spectrum Reusing Policy}
	Given any $\eta_{\mathit{EE}}$ in each iteration of $\mathbf{P1}$, our third-stage method is to finalized the optimal spectrum reusing policy for all CUEs and DUEs, based on the obtained power allocation solutions to both $\mathbf{P2}_{i,j}$ and $\mathbf{P3}_i$.
	First let $\lambda_{i,j}^*$ denote the maximum $\lambda_{i,j}$ at each potential spectrum reusing pair of CUE $i$ and DUE $j$ by solving each $\mathbf{P2}_{i,j}$, and let $\check{\lambda}_{i}^*$ denote the maximum $\check{\lambda}_{i}$ at the single CUE $i$ by solving each $\mathbf{P3}_i$.
	As such, the spectrum reusing problem becomes a variant of the weighted bipartite matching optimization problem, i.e.,
	\begin{align}
			\mathbf{P4}:\ \max_{\bm{\alpha}} \quad &  \sum_{i\in\mathcal{M}}\sum_{j\in\mathcal{N}}\alpha_{i,j}\lambda_{i,j}^*+\sum_{i\in\mathcal{M}}\check{\lambda}_{i}^*\left(1-\sum_{j\in\mathcal{N}}\alpha_{i,j}\right)~\label{P4}\\
			{\rm s.t.} \quad & \text{(\ref{P0}e)}-\text{(\ref{P0}g)}.\tag{\ref{P4}a}
	\end{align}
	Clearly, $\mathbf{P4}$ is an $M$-to-$N$ bipartite matching problem, which should be first expanded into an $M$-to-$M$ case for tractability as $M \geqslant N$.
	Specifically, let $\bm{\Omega}$ denote an $M \times M$ matrix, where all rows represent $M$ CUEs and the first $N$ columns represent all DUEs.
	In addition, the elements in the first $N$ columns of $\bm{\Omega}$ are filled with all $\lambda_{i,j}^*$, and the elements in each of the remaining $M-N$ columns are filled with the same $\check{\lambda}_{i}^*$.
	As such, $\bm{\Omega}$ is found by
	\begin{equation}
    \label{optimatrix}
    \bm{\Omega}=
    \overset{\text{Representing }N\text{ DUEs in } \mathcal{N}}{\left[
    \overbrace{\begin{array}{cccc:}
        \lambda_{1,1}^* & \lambda_{1,2}^* & \cdots & \lambda_{1,N}^*  \\
        \lambda_{2,1}^* & \lambda_{2,2}^* & \cdots & \lambda_{2,N}^* \\
        \vdots & \vdots & \ddots & \vdots  \\
        \lambda_{M,1}^* & \lambda_{M,2}^* & \cdots & \lambda_{M,N}^*  \\
    \end{array}}\right.}\!\!
    \overset{\text{Expanded }(M-N)\text{ columns}}{\left.\overbrace{\begin{array}{cccc}
        \check{\lambda}_{1}^* & \cdots & \check{\lambda}_{1}^*\\
        \check{\lambda}_{2}^* & \cdots & \check{\lambda}_{2}^* \\
        \vdots & \ddots & \vdots \\
        \check{\lambda}_{M}^* & \cdots & \check{\lambda}_{M}^*\\
    \end{array}}\right]}.
	\end{equation}
	
	Correspondingly, a new variable set needs to be defined as $\bm{\alpha}'=\left\{\alpha_{i,j'}\mid i \in \mathcal{M}, j' \in \mathcal{M}\right\}$, and then $\mathbf{P4}$ can be converted into an $M$-to-$M$ bipartite matching problem as
	\begin{align}
			\mathbf{P4.1}:\ \max_{\bm{\alpha}'} \quad &  \bm{\Omega}\odot \bm{\alpha}'~\label{P4.1}\\
			{\rm s.t.} \quad & \sum_{j'\in\mathcal{M}}\alpha_{i,j'}= 1,\ \forall i\in\mathcal{M},\tag{\ref{P4.1}a}\\
	& \sum_{i\in\mathcal{M}}\alpha_{i,j'}=1,\ \forall j' \in\mathcal{M},\tag{\ref{P4.1}b}\\
	& \alpha_{i,j'}\in \{0,1\}, \ \forall \left( i,j'\right) \in\mathcal{M}\times \mathcal{M},\tag{\ref{P4.1}c}
	\end{align}
	where the operator $\odot$ represents the Hadamard product.
	Note that the objective in $\mathbf{P4.1}$ remains scalar, and such a matrix representation is just to accommodate $\bm{\Omega}$ and $\bm{\alpha}'$.
	For better illustration, the expanded columns in $\bm{\Omega}$ can be interpreted as there hypothetically exist $(M-N)$ additional DUEs (corresponding to these expanded variables in $\bm{\alpha}'$) to divide up the remaining $(M-N)$ CUEs who are not sharing spectrum, but they have no impact on the energy efficiency gain of these CUEs.
	Hence, the problem transformation from $\mathbf{P4}$ to $\mathbf{P4.1}$ will not alter the original optimality and the optimal solution.
	
	Since $\mathbf{P4.1}$ is a standard bipartite matching problem, it can be efficiently solved in polynomial time by applying the Hungarian method~\cite{west2001introduction}.
	While $\mathbf{P4.1}$ can also be solved using the common relaxation and linear programming, it may lead to higher computational complexity and even suffer performance loss during the $0$-$1$ recovery process.
	After obtaining each optimal $\alpha_{i,j'}$ (denoted as $\alpha_{i,j'}^*$), it is easily derived that each DUE $j$ can reuse the spectrum of its optimal CUE $i$ (denoted by $\alpha_{i,j}^*$) if and only if it satisﬁes the following condition
	\begin{equation}
	\label{SR_determine}
		\alpha_{i,j}^*=\begin{cases}
			1,& \text{if } \alpha_{i,j'}^*=1 \text{ and } j' \in \mathcal{N}=\{1,2,\cdots,N\};\\
			0,& \text{otherwise}.
		\end{cases}
	\end{equation}
	In this way, given any $\eta_{\mathit{EE}}$ in each iteration w.r.t. $\mathbf{P1}$, the optimal spectrum reusing scheme is now finalized along with the optimal power allocation policy at each CUE and DUE.
	In practice, the BS can act as a network controller, centrally optimizing and assigning the spectrum and power resources in D2D-SCNs based on the efficient solution designed above.
	
	
	\subsection{Algorithm Analysis}
	To better demonstrate the full procedure of the proposed resource allocation solution in Fig.~\ref{Solution}, we summarize its key technical points and enclose them in Algorithm~\ref{Algo1}, which is placed in the right column on this page.
	\begin{algorithm}[htbp]
				\caption{Proposed Resource Allocation for D2D-SCN}
				\label{Algo1}
				\begin{algorithmic}[1]
					\REQUIRE \textit{The network parameters $M$, $N$, $K$, $W$, $L$, $\delta^{2}$, $P^{\mathit{enc}}$, $\xi$, $P_{\mathit{max}}^{C}$, $P_{\mathit{max}}^{D}$, $V_{min}^{C}$, $V_{min}^{D}$, and each SemCom user's parameters $G_{i, B}$, $G_{j, B}$, $G_{j}^{D}$, $G_{i,j}$, $u_{i,k}^{C}$, $u_{j,k}^{D}$, $\beta_{i}^{C}$, $\beta_{j}^{D}$}
					\ENSURE \textit{The power allocation policy $\left(\bm{P^{C^{*}}},\bm{P^{D^{*}}}\right)$ and the spectrum reusing strategy $\bm{\alpha}^{*}$}
					\STATE \textit{Initialize iteration index $t\leftarrow1$ and $\eta_{\mathit{EE}}(1)\leftarrow0$ for $\mathbf{P1}$}
					\STATE \textit{Set $Q$ (large) and $\epsilon$ (small) to proper positive values}
					\WHILE{$t\leqslant Q$}
						\FOR{$i \leftarrow1$ \textit{to} $M$}
							\FOR{$j \leftarrow1$ \textit{to} $N$}
								\STATE \textit{Initialize the search round index as $\tilde{t}\leftarrow1$ and the
								\STATE \quad solution list $\mathcal{I}(1)\leftarrow\emptyset$ for solving each $\mathbf{P2}_{i,j}$
								\STATE Set the maximum number of search rounds as $\tilde{Q}$}
								\WHILE{$\tilde{t}\leqslant \tilde{Q}$}
									\STATE \textit{Choose an arbitrary point in $\psi$ as $\left(\widetilde{P_{i}^{C}},\widetilde{P_{j}^{D}}\right)$}
									\STATE \textit{Find $\left(\overleftarrow{P_{i}^{C}},\overleftarrow{P_{j}^{D}}\right)$ w.r.t. $\left(\widetilde{P_{i}^{C}},\widetilde{P_{j}^{D}}\right)$ by Phase (2)}
									\STATE \textit{Find $\left(\overrightarrow{P_{i}^{C}},\overrightarrow{P_{j}^{D}}\right)$ w.r.t. $\left(\overleftarrow{P_{i}^{C}},\overleftarrow{P_{j}^{D}}\right)$ by Phase (3)}
									\WHILE{$\left(\overleftarrow{P_{i}^{C}},\overleftarrow{P_{j}^{D}}\right)\neq\left(\overrightarrow{P_{i}^{C}},\overrightarrow{P_{j}^{D}}\right)$}
										\STATE \textit{$\left(\widetilde{P_{i}^{C}},\widetilde{P_{j}^{D}}\right)\leftarrow\left(\overrightarrow{P_{i}^{C}},\overrightarrow{P_{j}^{D}}\right)$}
										\STATE \textbf{repeat} \textit{Lines 11 and 12}
									\ENDWHILE
									\STATE \textit{$\mathcal{I}\left(\tilde{t}+1\right)\leftarrow\mathcal{I}\left(\tilde{t}\right)\cup\left\{\left(\overleftarrow{P_{i}^{C}},\overleftarrow{P_{j}^{D}}\right)\right\}$}
									\STATE \textit{$\tilde{t} \leftarrow \tilde{t} +1$}
								\ENDWHILE
								\STATE \textit{Finalize $\left(P_{i}^{C^*},P_{j}^{D^*}\right)$ for each potential CUE $i$-}
								\STATE \quad \textit{DUE $j$ pair by comparing all solutions in $\mathcal{I}(\tilde{Q})$}
								\STATE \textit{Compute $\lambda_{i,j}^*(t)$ by (\ref{P2}) for $\mathbf{P4}$}
							\ENDFOR
						\ENDFOR
						\FOR{$i \leftarrow1$ \textit{to} $M$}
							\STATE \textit{Finalize the optimal $P_{i}^{C}$ for each single CUE $i$ by}
							\STATE \quad \textit{employing the brute-force search to solve $\mathbf{P3}_i$}
							\STATE \textit{Compute $\check{\lambda}_{i}^*(t)$ by (\ref{P3}) for $\mathbf{P4}$}
						\ENDFOR
						\STATE \textit{Generate $\bm{\Omega}$ according to \eqref{optimatrix}}
						\STATE \textit{Solve $\mathbf{P4.1}$ by using the Hungarian method}
						\STATE \textit{Determine each $\alpha_{i,j}^{*}(t)$ by \eqref{SR_determine}}
					\STATE \textit{Finalize $\bm{P^{C^{*}}}(t)$, $\bm{P^{D^{*}}}(t)$, and $\bm{\alpha^{*}}(t)$ by backtracking}
					\STATE \quad \textit{through the element corresponding to $\alpha_{i,j}^{*}(t)$ in $\bm{\Omega}$}
					\STATE \textit{Calculate $F\left(\eta_{\mathit{EE}}(t)\right)$ by substituting the finalized}
					\STATE \quad \textit{$\bm{P^{C^{*}}}(t)$, $\bm{P^{D^{*}}}(t)$, and $\bm{\alpha^{*}}(t)$ into (\ref{P1})}
					\IF{$F\left(\eta_{\mathit{EE}}(t)\right)<\epsilon$}
						\RETURN $\left(\!\bm{P^{C^{*}}}\!,\bm{P^{D^{*}}}\!, \bm{\alpha}^{*}\right)\!\leftarrow\!\left(\!\bm{P^{C^{*}}}\!(t),\bm{P^{D^{*}}}\!(t), \bm{\alpha^{*}}\!(t)\right)$
						\STATE \textbf{break}
					\ELSE
						\STATE \textit{Update $\eta_{\mathit{EE}}(t+1)$ by \eqref{update}}
						\STATE $t\leftarrow t+1$
					\ENDIF
					\ENDWHILE
					\RETURN $\left(\!\bm{P^{C^{*}}}\!,\bm{P^{D^{*}}}\!, \bm{\alpha}^{*}\right)\!\leftarrow\!\left(\!\bm{P^{C^{*}}}\!(Q),\bm{P^{D^{*}}}\!(Q), \bm{\alpha^{*}}\!(Q)\right)$
			\end{algorithmic}
		\end{algorithm}
	
	Regarding the computational complexity of Algorithm~\ref{Algo1}, it is first seen that in each search round for solving each $\mathbf{P2}_{i,j}$, it takes several iterations (Lines 11-16) to determine one viable solution in $\mathcal{I}$, and in each iteration, the brute-force search needs to be executed once to obtain $\left(\overleftarrow{P_{i}^{C}},\overleftarrow{P_{j}^{D}}\right)$ or $\left(\overrightarrow{P_{i}^{C}},\overrightarrow{P_{j}^{D}}\right)$.
	Hence, if denoting the maximum number of iterations w.r.t. Lines 11-16 as $H$ and the maximum number of staircase w.r.t. each brute-force search as $\tilde{H}$, then solving each $\mathbf{P2}_{i,j}$ would have the complexity of $\mathcal{O}(\tilde{Q}H\tilde{H})$.
	Likewise, solving each $\mathbf{P3}_i$ has the complexity of $\mathcal{O}(\tilde{H})$.
	Moreover, since the $M$-to-$M$ bipartite matching problem w.r.t. $\mathbf{P4.1}$ can be solved by the Hungarian method with the complexity of $\mathcal{O}(M^3)$~\cite{edmonds1972theoretical}, the complexity for solving $\mathbf{P4}$ is $\mathcal{O}(M^3)$.
	Accordingly, the proposed Algorithm~\ref{Algo1} has a polynomial-time overall complexity of $\mathcal{O}(QMN\tilde{Q}H\tilde{H}+M^3)$.
	
	\section{Numerical Results and Discussions}
	In this section, numerical evaluations are conducted to demonstrate the performance of our proposed power allocation and spectrum reusing solutions in the D2D-SCN, where we employ Python 3.7-based PyCharm as the simulator platform and implement it in a workstation PC featuring the AMD Ryzen-9-7900X processor with 12 CPU cores and 128 GB RAM.
	In the basic system setup, we first model a single-cell circular area with a radius of $300$ meters, in which multiple CUEs and DUEs are randomly dropped, and the distance between the transceiver of each DUE is randomly generated between $50$ and $200$ meters.
    For the energy efficiency model, each SemCom device is assumed to have a fixed power amplifier efficiency of $35$ percent~\cite{wang2006realistic}, i.e., $\xi=1/0.35=2.8571$, and let the circuit power consumption required for encoding one semantic triplet $P^{\mathit{enc}}=0.5$ mW~\cite{liew2022economics}.
    Note that no DL-related simulations are involved here and $P^{enc}$ is incorporated as an average power cost per semantic triplet based on existing measurements, since the focus of this work is dedicatedly on evaluating the proposed system-level resource allocation solution, rather than encoder design.
    For brevity, other simulation parameters along with their values are summarized in Table~\ref{SimuPara}.
    \begin{table}[t]
		\centering
		\caption{Simulation Parameters}
		\label{SimuPara}
		\setlength{\tabcolsep}{3pt}
		\renewcommand\arraystretch{1.5}
		\begin{tabular}{|m{4.2cm}<{\raggedright}|m{3.8cm}<{\raggedright}|}\hline
			\textbf{Parameters} & \textbf{Values} \\ \hline
			Number of SemCom-enabled CUEs ($M$) & $50$\\ \hline
			Number of SemCom-enabled DUEs ($N$) & $30$\\ \hline
			System bandwidth & $10$ MHz\\ \hline
			Maximum transmit power of each CUE ($P_{max}^{C}$) & $23$ dBm~\cite{pawar2021joint} \\ \hline
			Maximum transmit power of each DUE ($P_{max}^{D}$) & $21$ dBm~\cite{pawar2021joint} \\ \hline
			Noise power ($\delta^{2}$) & $-111.45$ dBm \\ \hline
			Path loss model for cellular links & $128.1+37.6\log_{10}\left(d\ \text{[km]}\right)$ dB\\ \hline
			Path loss model for D2D links & $148+40\log_{10}\left(d\ \text{[km]}\right)$ dB~\cite{esmat2018uplink}\\ \hline
			Average number of bits required for encoding one semantic triplet ($L$) & $50$ bits~\cite{gao2024importance}\\ \hline
		\end{tabular}
	\end{table}
	
    For the simulated SemCom model, we set $K=20$ as the total number of wireless SemCom services in the D2D-SCN.
    Moreover, the preference ranking of each CUE (i.e., $u_{i,k}^{C}$) and DUE (i.e., $u_{j,k}^{D}$) for all these services is generated in an independent and random manner, as different sorting ways of them do not affect its total semantic value over all SemCom services.
    The skewness of the Zipf distribution at each CUE (i.e., $\beta_{i}^{C}$) and DUE (i.e., $\beta_{j}^{D}$) is randomly distributed in a range of $0.5\sim 1.5$~\cite{gao2024importance}.
	As for the solution settings, the maximum number of iterations for updating $\eta_{\mathit{EE}}$ in $\mathbf{P1}$ is set as $Q=20$, and its convergence threshold $\epsilon$ is $0.01$.
	In addition, the minimum semantic value threshold is set as the same for all CUEs and DUEs, i.e., $V_{min}=V_{min}^{C}=V_{min}^{D}=50$.
	It is worth mentioning that all the above parameter values are set by default unless otherwise specified, and all subsequent numerical results are obtained by averaging over a sufficiently large number of trials.
	
	For comparison purposes, here we employ two resource allocation benchmarks in D2D-SCNs: (I) Maximum power allocation plus random spectrum reusing~\cite{guo2019resource}, which means that each user is allocated with its maximum allowable transmit power while each DUE randomly reuses the subchannel of one CUE; (II) Random power allocation~\cite{kim2005random} plus distance-based spectrum reusing~\cite{ningombam2018non}, where each user is allocated with the randomized transmit power while each DUE reuses the subchannel of the CUE furthest away from itself to reduce the interference impact as much as possible.\footnote{The two baselines are considered adequate for comparison due to their general and adaptable purposes. Additional competitive baselines will be incorporated into future extensions for further validation and improvement.}
	
\begin{figure}[t]
		\centering
		\includegraphics[width=0.45\textwidth]{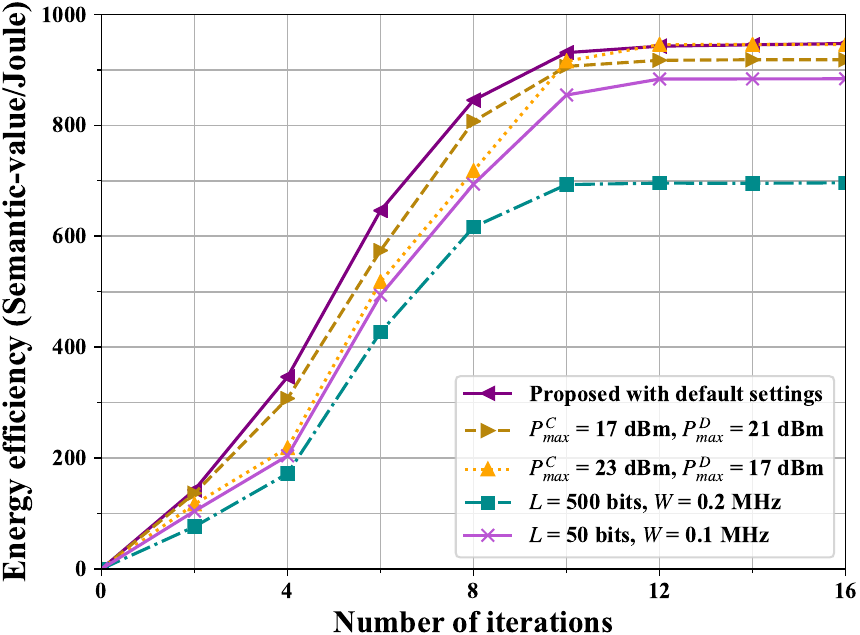} 
		\caption{Energy efficiency ($\eta_{\mathit{EE}}$) versus different numbers of iterations.}
		\label{PowerFigure0}
    \end{figure}

    Fig.~\ref{PowerFigure0} first shows the energy efficiency performance (i.e., $\eta_{\mathit{EE}}$)  with different numbers of iterations $Q$ to verify the convergence and parameter sensitivity of the proposed solution, where different settings of $P_{max}^{C}=17$ dBm, $P_{max}^{D}= 17$ dBm, $L=500$ bits, and $W=0.1$ MHz are used in comparison with those in Table~\ref{SimuPara}.
    With the increase of iterations, it can be observed that the proposed solution in different settings can always converge to their respective maximum $\eta_{\mathit{EE}}$ while having the same convergence speed, i.e., they all reach convergence after around $12$ iterations.
    This phenomenon indicates that our designed algorithm is not affected by changes of system parameter values and is therefore very ideal for practical implementation.
    Meanwhile, compared with the default settings, Fig.~\ref{PowerFigure0} further shows that the worse $\eta_{\mathit{EE}}$ occurs at the smaller $P_{max}^{C}$ and $W$ as well as the larger $L$.
	The underlying reason is that the stronger limitation on CUEs' maximum transmit power leads to a certain degree of semantic value degradation especially for those CUEs without spectrum sharing, as they tend to use the maximum $P_{max}^{C}$ compared with these with spectrum sharing.
	In addition, both the smaller $W$ and the larger $L$ can result in the lower achievable rate of semantic triplet transmission per link, which justifies the reduction in the energy efficiency.
	
	\begin{figure}[t]
		\centering
		\includegraphics[width=0.45\textwidth]{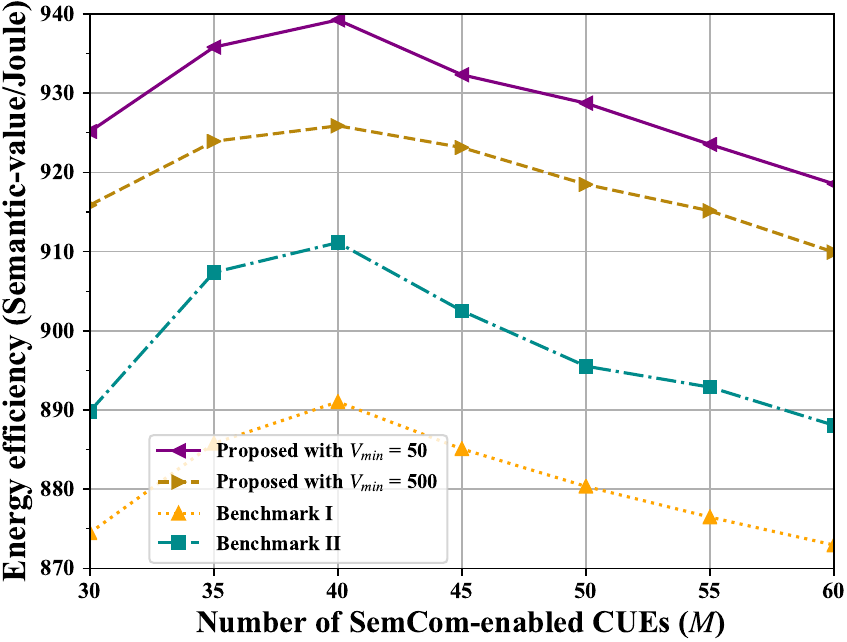} 
		\caption{Energy efficiency versus different numbers of CUEs.}
		\label{PowerFigure1}
    \end{figure}
     Fig.~\ref{PowerFigure1} shows the objective performance metric of $\eta_{\mathit{EE}}$ obtained under varying numbers of CUEs (i.e., $M$) between $30$ and $60$, where two different semantic value thresholds of $V_{min}=50$ and $V_{min}=500$ are considered.
    Compared with the two benchmarks, it is seen that our proposed solution always guarantees a significant performance gain with the changes of $M$.
    For instance, when $M=35$, the energy efficiency of $935.8$ semantic-value/Joule is observed by the proposed solution at $V_{min}=50$, which increases $5.76\%$ performance compared to Benchmark I and $3.2\%$ compared to Benchmark II.
    Moreover, as $M$ increases, the energy efficiency of the proposed solution rises at the beginning from $30$ to $40$, and then drops gradually.
    This is because the increase of $M$ at the beginning can provide each DUE more options to choose a better CUE for spectrum reusing and thus leading to a better energy efficiency performance.
    Then, as $M$ keeps growing, the subchannel bandwidth $W$ averaged to each CUE becomes fewer due to the fixed total system bandwidth, but the number of DUEs remains constant in Fig.~\ref{PowerFigure1}.
    Hence, after the point of $M=40$, the reduction in energy efficiency due to less allocated bandwidth dominates the trend of $\eta_{\mathit{EE}}$, even though each DUE can still have a better spectrum reusing option.
    In addition, the proposed solution with $V_{min}=500$ has a worse $\eta_{\mathit{EE}}$ performance compared to that with $V_{min}=50$.
	This can be explained by that $V_{min}=500$ represents the stringent constraints of (\ref{P0}a) and (\ref{P0}b) and thus results in a smaller feasible region of variables $\bm{P^{C}}$ and $\bm{P^{D}}$ in $\mathbf{P0}$, which incurs a deterioration of $\eta_{\mathit{EE}}$ that can be achieved.
    
    \begin{figure}[t]
		\centering
		\includegraphics[width=0.45\textwidth]{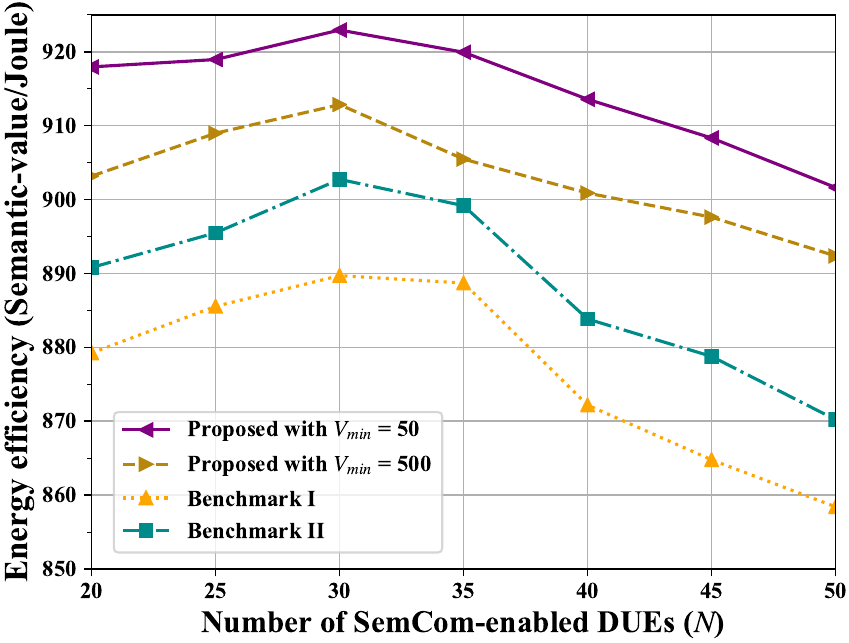} 
		\caption{Energy efficiency versus different numbers of DUEs.}
		\label{PowerFigure2}
    \end{figure}
    Next, we compare the proposed solution with the two benchmarks under varying number of DUEs (i.e., $N$) between $20$ and $50$ at the same two semantic value thresholds in Fig.~\ref{PowerFigure2}.
    It is observed that the energy efficiency $\eta_{\mathit{EE}}$ obtained by our solution still exceeds the benchmarks at each point with a significant performance gain.
    Likewise, $\eta_{\mathit{EE}}$ becomes higher with $N$ at the beginning, and then decreases when exceeding $30$.
    The former phenomenon is because the performance gain on $V^{\mathit{total}}$ resulted from the increase of $N$ surpasses the impact of the power consumption increase on $E^{\mathit{total}}$.
    When $N$ surpasses a maximum threshold (i.e., $N=40$ in our case), such performance increase eventually reaches its peak and is saturated and even worse.
    This occurs because interference between spectrum-sharing links becomes dominant, leading to a decrease in $\eta_{\mathit{EE}}$, as more CUEs share spectrum with a larger number of DUEs.
    Moreover, the higher number of DUEs indicates the more restrictive spectrum sharing optimization, which also brings a certain degree of performance limitation.
    
    \begin{figure}[t]
		\centering
		\includegraphics[width=0.45\textwidth]{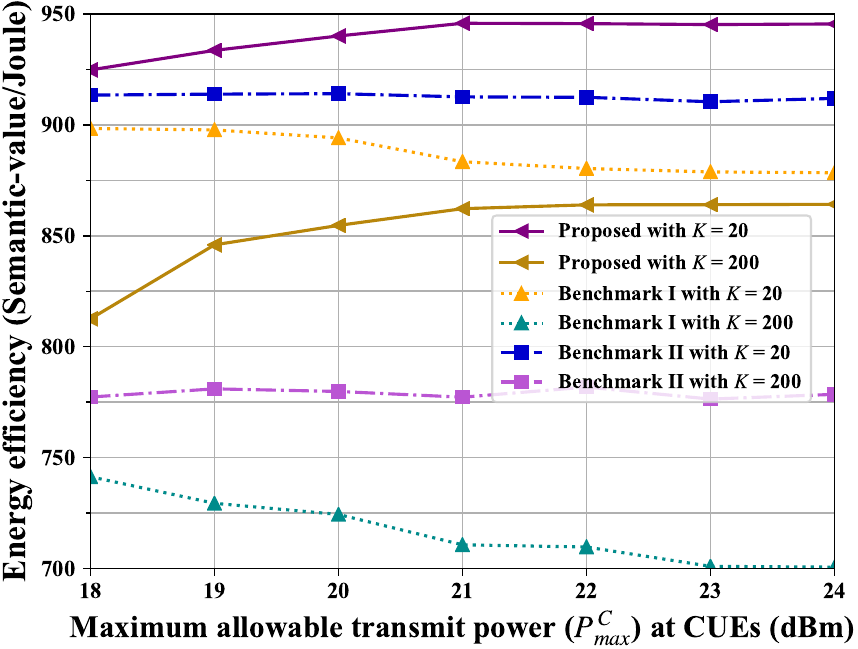} 
		\caption{Energy efficiency versus varying maximum transmit power of CUEs.}
		\label{PowerFigure3}
    \end{figure}
\begin{figure}[t]
		\centering
		\includegraphics[width=0.45\textwidth]{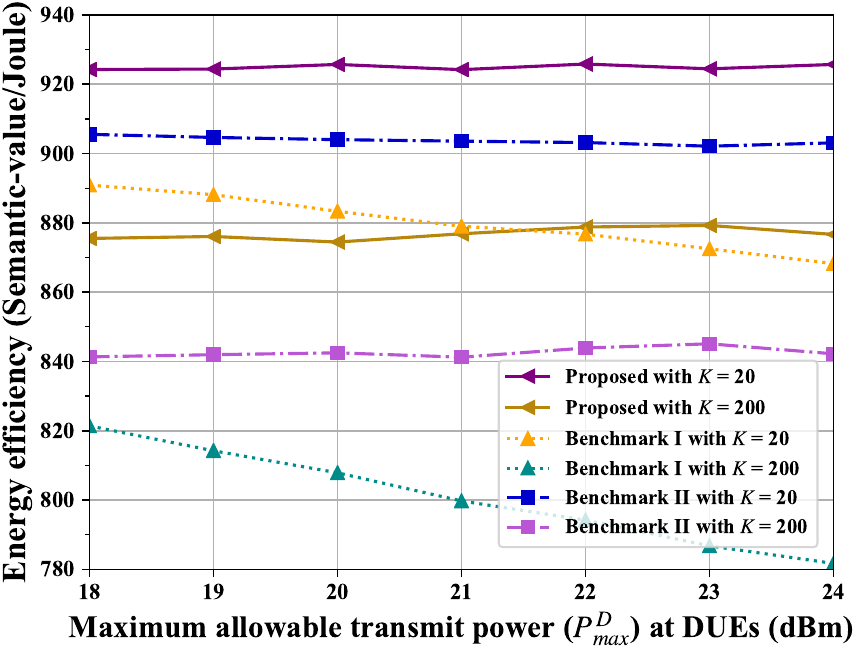} 
		\caption{Energy efficiency versus varying maximum transmit power of DUEs.}
		\label{PowerFigure4}
    \end{figure}

    Apart from these, the impacts of the maximum allowable transmit power of CUEs ($i.e., P_{max}^{C}$) and DUEs (i.e., $P_{max}^{D}$) are tested in Fig.~\ref{PowerFigure3} and Fig.~\ref{PowerFigure4}, respectively, where two differing numbers of SemCom services $K=20$ and $K=200$ are taken into account.
    First, it is seen in both figures that our proposed solution renders a better energy efficiency performance compared with the two benchmarks with the same $K$.
    In Fig.~\ref{PowerFigure3}, as $P_{max}^{C}$ rises from $18$ to $24$ dBm, an increase is observed by the proposed solution at both $K$ curves and then eventually stabilized.
    This can be understood by that CUEs prefer higher transmit power at the beginning under looser power constraints to improve semantic-value performance, particularly for CUEs without spectrum reuse.
    Afterward, when $P_{max}^{C}$ keeps increasing, the optimal transmit power for these CUEs tends to be stable, as the further increasing of power will lead to more energy consumption and the worse $\eta_{\mathit{EE}}$ performance.
    In addition, Fig.~\ref{PowerFigure4} depicts a steady trend as $P_{max}^{D}$ grows when compared with Fig.~\ref{PowerFigure3} in the performance presentation of our solution.
    This is because the optimal transmit power for DUEs in our proposed solution should be in a small region to align with the corresponding minimum semantic value constraint $V_{min}$, and therefore, the increase of $P_{max}^{D}$ will not affect the final $\eta_{\mathit{EE}}$ performance.
    Specially, the comparison of Fig.~\ref{PowerFigure3} and Fig.~\ref{PowerFigure4} demonstrates a fact that $\eta_{\mathit{EE}}$ should be more sensitive to CUEs' transmit power cap than DUEs' one.
    While both power budgets matter in the D2D-SCN, the CUEs' power cap is the primary bottleneck for semantic-level energy efficiency.
    
    Moreover, in either Fig.~\ref{PowerFigure3} or~\ref{PowerFigure4}, Benchmark I always shows a downtrend while Benchmark II shows a stable and unchanging trend.
    The former is because Benchmark I is based on the maximum transmit power assignment that can result in significant power dissipation $E^{T}$, and thus leading to a downtrend of $\eta_{\mathit{EE}}$.
    The latter phenomenon is due to the Benchmark II's random power allocation policy, which will not obviously affect the transmit power assigned to each SemCom user.
    In fact, the two benchmarks either neglect semantic utility or use non-targeted power allocation patterns, leading to less adaptive behavior and poor performance, however, the proposed solution dynamically adapts to the changes in both user density and transmit power constraints.
    Furthermore, $\eta_{\mathit{EE}}$ at each of the three schemes with the smaller $K$ is always higher than that with the larger one.
    This is due to the fact that a larger $K$ implies less discrepancy in users' preferences for different SemCom services given the fixed skewness of Zipf distribution, and these services with low semantic value will become inevitably dominant, thereby causing the worse energy efficiency.
    
    \begin{figure}[t]
		\centering
		\includegraphics[width=0.45\textwidth]{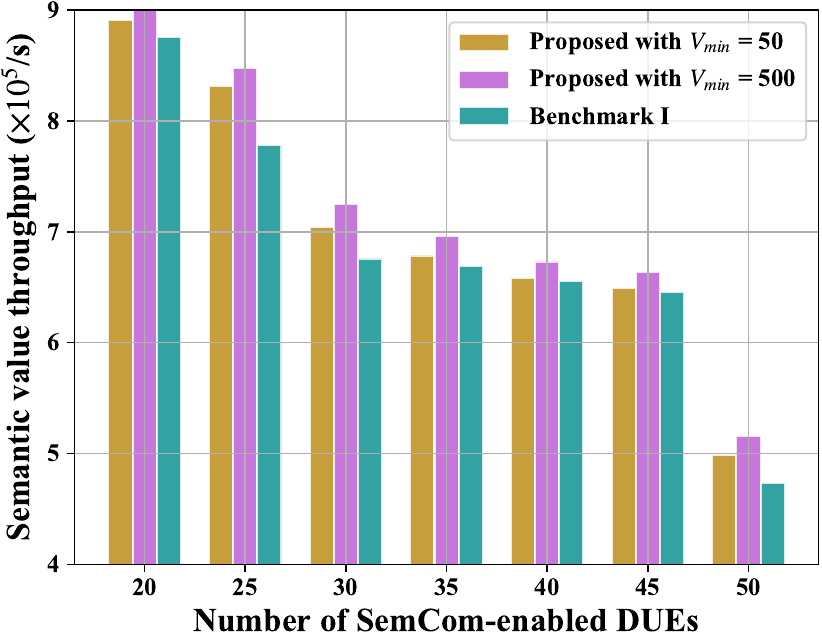} 
		\caption{Semantic value throughput ($V^{\mathit{total}}$) versus different numbers of DUEs.}
		\label{PowerFigure5}
    \end{figure}
     \begin{figure}[t]
		\centering
		\includegraphics[width=0.45\textwidth]{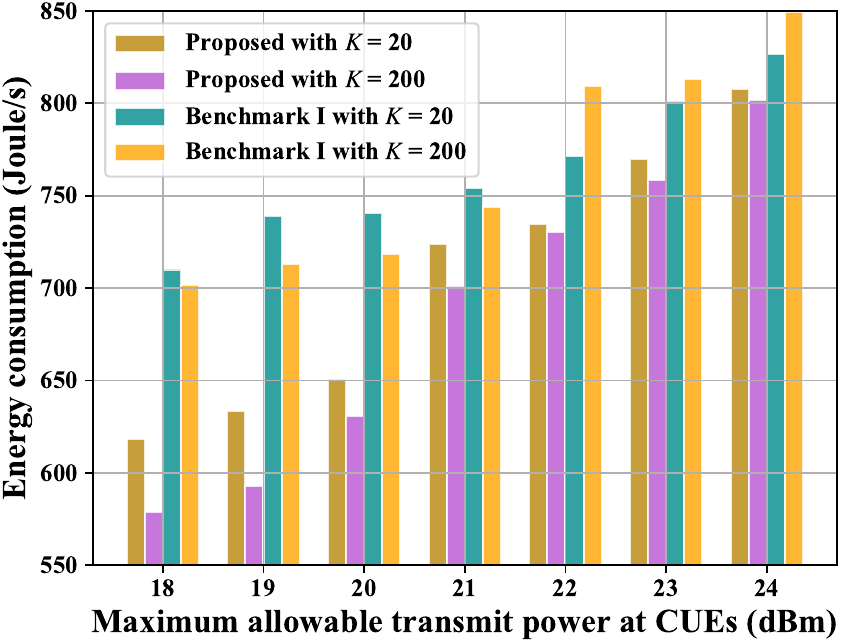} 
		\caption{Overall energy consumption ($E^{\mathit{total}}$) versus varying $P_{max}^{C}$.}
		\label{PowerFigure6}
    \end{figure}
    Finally, Fig.~\ref{PowerFigure5} and Fig.~\ref{PowerFigure6} demonstrate the semantic value throughput (i.e., $V^{\mathit{total}}$) under varying $N$ with different $V_{min}$ and the overall energy consumption (i.e., $E^{\mathit{total}}$) under varying $P_{max}^{C}$ with different $K$, respectively.
    It is first seen in Fig.~\ref{PowerFigure5} that $V^{\mathit{total}}$ gradually drops as the increase of $N$, and the setting of $V_{min}=500$ leads to a better performance than $V_{min}=50$.
    The former trend is because that the increase of $N$ makes more CUEs share their spectrum that can cause greater signal interference and worse SINR to them, thus resulting in the decrease of overall bit throughout as well as overall semantic value.
    As for the latter phenomenon, this is relatively clear since $V_{min}=500$ sets a higher semantic value threshold for each CUE and DUE link that must be satisfied during our optimization process.
    Moreover, the semantic value observed by our solution always outperforms benchmark I at each point.
    This can be interpreted by that the benchmark I employs the maximum power allocation scheme, bringing very heavy interference burden on these CUEs and DUEs with spectrum sharing, which performance degradation outweighs the performance gain from the maximum power to those CUEs without spectrum sharing.
    As plotted in Fig.~\ref{PowerFigure6}, an uptrend of $E^{\mathit{total}}$ is observed for all resource allocation schemes as the growing of $P_{max}^{C}$.
    This result is expected because for these CUEs with no spectrum sharing, they prefer to perform SemCom with their maximum allowable transmit power to maximize the achievable semantic value performance, since they have no concerns for signal interferences compared with these CUEs with spectrum sharing.
   It is noticed that the proposed solution with the larger $K$ consumes less energy than that with the small $K$.
   This can be understood by that the larger $K$ means the larger $\theta_{i}^{C}$ and $\theta_{j}^{D}$ in~(\ref{cueSV}) and~(\ref{dueSV}) from the statistical perspective, which represents the looser constraints for (\ref{P0}a) and (\ref{P0}b).
   As such, the smaller transmit power may be utilized by CUEs and DUEs to meet their minimum semantic value threshold, and thus leading to the less energy consumption.
   Again, the proposed solution achieves less energy consumption in comparison with benchmark I at each $P_{max}^{C}$ setting for the same $K$, which is consistent with the previous results in Fig.~\ref{PowerFigure3} and proves the performance superiority of the proposed solution.
   It is concluded that our resource allocation strategy fully considers user-specific preferences and semantic encoding power consumption, decoupling energy efficiency from conventional bit-centric paradigms and thereby focusing on improving the quality of semantic service experiences.
	
	\section{Conclusions}
	In this paper, we have investigated the wireless resource allocation problem for energy efficiency maximization in the novel network scenario of D2D-SCN, where multiple CUEs and DUEs coexist for SemCom service provisioning.
	To measure the semantic information importance in SemCom, the user preference-aware semantic triplet transmission has been first introduced to identify the semantic value performance, while taking into account the power circuit consumption from the specific semantic encoding mechanism for energy efficiency.
	On this basis, a joint power control and spectrum reuse problem has been then formulated to maximize the energy efficiency of D2D-SCN.
	After the fractional-to-subtractive primal problem transformation and decomposition, we have developed a three-stage method to seek the optimal resource allocation solution with low computational complexity, and the solution optimality has been theoretically proved.
	Numerical results have verified the performance superiority of the proposed solution in terms of energy efficiency, semantic value, and energy consumption compared with differing benchmarks.
	
	This work can serve as a pioneer in providing valuable insights for follow-up research on D2D SemCom underlying cellular networks.
	Other relevant networking issues in the D2D-SCN, such as cellular/D2D SemCom mode switching and semantic security-driven or user fairness-aware resource allocation, can treat this paper as the theoretical baseline for reference.
	Moreover, since this work is well-suited to structured semantic representations for SemCom, a further extension on the semantic value measurement for low-level semantic reconstruction tasks, where semantics are not symbolically discrete, could be our next research direction.
	In addition, the tradeoffs between communication and computing loads and the joint source-and-channel coding-based semantic value measurement are also worth exploring in the future.
	
	\begin{appendices}
		\section{Proof of Proposition 1}
		First let $\left(\bm{P^{C^{*}}},\bm{P^{D^{*}}}, \bm{\alpha^{*}}\right)$ and $\left(\bm{\widehat{P^{C}}},\bm{\widehat{P^{D}}}, \bm{\widehat{\alpha}}\right)$ denote the optimal solution and an arbitrary feasible solution to $\mathbf{P0}$, respectively, and let $\eta_{\mathit{EE}}^{*}$ denote the optimal value that can be reached in $\mathbf{P0}$.
		Clearly, we have
		\begin{equation}
		\label{prop1eqa}
		\begin{aligned}
			\eta_{\mathit{EE}}^{*}=\frac{V^{\mathit{total}}\left(\bm{P^{C^{*}}},\bm{P^{D^{*}}}, \bm{\alpha^{*}}\right)}{E^{\mathit{total}}\left(\bm{P^{C^{*}}},\bm{P^{D^{*}}}, \bm{\alpha^{*}}\right)}\geqslant \frac{V^{\mathit{total}}\left(\bm{\widehat{P^{C}}},\bm{\widehat{P^{D}}}, \bm{\widehat{\alpha}}\right)}{E^{\mathit{total}}\left(\bm{\widehat{P^{C}}},\bm{\widehat{P^{D}}}, \bm{\widehat{\alpha}}\right)}.
		\end{aligned}
		\end{equation}
	Since $E^{\mathit{total}}>0$ holds for any feasible solution, (\ref{prop1eqa}) implies
	\begin{equation}
		\label{prop1eqb}
		\begin{aligned}
			V^{\mathit{total}}\left(\bm{P^{C^{*}}},\bm{P^{D^{*}}}, \bm{\alpha^{*}}\right)\!-\!\eta_{\mathit{EE}}^{*}E^{\mathit{total}}\left(\bm{P^{C^{*}}},\bm{P^{D^{*}}}, \bm{\alpha^{*}}\right)=0,
		\end{aligned}
		\end{equation}
		and
		\begin{equation}
		\label{prop1eqc}
		\begin{aligned}
			V^{\mathit{total}}\left(\bm{\widehat{P^{C}}},\bm{\widehat{P^{D}}}, \bm{\widehat{\alpha}}\right)\!-\!\eta_{\mathit{EE}}^{*}E^{\mathit{total}}\left(\bm{\widehat{P^{C}}},\bm{\widehat{P^{D}}}, \bm{\widehat{\alpha}}\right)\leqslant 0.
		\end{aligned}
		\end{equation}
			
		Note that $\mathbf{P0}$ and $\mathbf{P1}$ have the same feasible region as they have the identical constraints (\ref{P0}a)-(\ref{P0}g).
		If substituting $\eta_{\mathit{EE}}^{*}$ into the objective function of $\mathbf{P1}$, it is obviously seen from \eqref{prop1eqb} and \eqref{prop1eqc} that $\mathbf{P1}$'s achievable maximum objective value, i.e., $F\left(\eta_{\mathit{EE}}^{*}\right)$, is also $0$, while its optimal solution is exactly the optimal solution to $\mathbf{P0}$, i.e., $\left(\bm{P^{C^{*}}},\bm{P^{D^{*}}}, \bm{\alpha^{*}}\right)$.
		
		Meanwhile, considering the other two remaining cases of $F\left(\eta_{\mathit{EE}}\right)>0$ and $F\left(\eta_{\mathit{EE}}\right)<0$ in $\mathbf{P1}$, and let $\left(\overline{\bm{P^{C^{*}}}},\overline{\bm{P^{D^{*}}}}, \overline{\bm{\alpha^{*}}}\right)$ and $\left(\widetilde{\bm{P^{C^{*}}}},\widetilde{\bm{P^{D^{*}}}}, \widetilde{\bm{\alpha^{*}}}\right)$ be their optimal solutions to $\mathbf{P1}$ respectively, then we have
		\begin{equation}
		\label{prop1eqe}
		\begin{aligned}
			V^{\mathit{total}}\left(\overline{\bm{P^{C^{*}}}},\overline{\bm{P^{D^{*}}}}, \overline{\bm{\alpha^{*}}}\right)\!-\!\eta_{\mathit{EE}}E^{\mathit{total}}\left(\overline{\bm{P^{C^{*}}}},\overline{\bm{P^{D^{*}}}}, \overline{\bm{\alpha^{*}}}\right)>0,
		\end{aligned}
		\end{equation}
		and
		\begin{equation}
		\label{prop1eqf}
		\begin{aligned}
			V^{\mathit{total}}\left(\widetilde{\bm{P^{C^{*}}}},\widetilde{\bm{P^{D^{*}}}}, \widetilde{\bm{\alpha^{*}}}\right)\!-\!\eta_{\mathit{EE}}E^{\mathit{total}}\left(\widetilde{\bm{P^{C^{*}}}},\widetilde{\bm{P^{D^{*}}}}, \widetilde{\bm{\alpha^{*}}}\right)<0.
		\end{aligned}
		\end{equation}
		Again leveraging $E^{\mathit{total}}>0$, given any $\eta_{\mathit{EE}}$, (\ref{prop1eqe}) yields
		\begin{equation}
		\label{prop1eqg}
		\begin{aligned}
			\frac{V^{\mathit{total}}}{E^{\mathit{total}}}=\eta_{\mathit{EE}}<\frac{V^{\mathit{total}}\left(\overline{\bm{P^{C^{*}}}},\overline{\bm{P^{D^{*}}}}, \overline{\bm{\alpha}^{*}}\right)}{E^{\mathit{total}}\left(\overline{\bm{P^{C^{*}}}},\overline{\bm{P^{D^{*}}}}, \overline{\bm{\alpha^{*}}}\right)},
		\end{aligned}
		\end{equation}
		and likewise, (\ref{prop1eqf}) yields
		\begin{equation}
		\label{prop1eqh}
		\begin{aligned}
			\frac{V^{\mathit{total}}}{E^{\mathit{total}}}=\eta_{\mathit{EE}}>\frac{V^{\mathit{total}}\left(\widetilde{\bm{P^{C^{*}}}},\widetilde{\bm{P^{D^{*}}}}, \widetilde{\bm{\alpha^{*}}}\right)}{E^{\mathit{total}}\left(\widetilde{\bm{P^{C^{*}}}},\widetilde{\bm{P^{D^{*}}}}, \widetilde{\bm{\alpha^{*}}}\right)}.
		\end{aligned}
		\end{equation}
		From (\ref{prop1eqg}) and (\ref{prop1eqh}), it can be concluded that for any feasible solution to $\mathbf{P0}$, it must not be the optimal solution to $\mathbf{P1}$ when $F\left(\eta_{\mathit{EE}}^{*}\right)\neq0$.
		This completes the proof.
			
		\section{Proof of Proposition 2}
		This proposition can be proved by using contradiction.
		According to all constraints of $\mathbf{P2}_{i,j}$, if the optimization problem is supposed to be solvable, the optimal power allocation solution $\left(P_{i}^{C^*},P_{j}^{D^*}\right)$ must fall into the non-empty $\psi$.
		Here, we first assume that $\left(P_{i}^{C^*},P_{j}^{D^*}\right)$ is not the coincide point of the two line segments while making $\widetilde{\lambda_{\mathit{1}}}\left(P_{j}^{D}\right)$ and $\widetilde{\lambda_{\mathit{2}}}\left(P_{i}^{C}\right)$ simultaneously reach their respective maxima, i.e.,
		\begin{equation}
			\left(P_{i}^{C^*}\!,P_{j}^{D^*}\!\right)\!\notin\! \left\{\!\left(\overleftarrow{P_{i}^{C}},\overleftarrow{P_{j}^{D}}\right)\!\mid\! \left(\overleftarrow{P_{i}^{C}},\overleftarrow{P_{j}^{D}}\right)\!=\!\left(\overrightarrow{P_{i}^{C}}\!,\overrightarrow{P_{j}^{D}}\right)\!\in \!\psi\right\}.
		\end{equation}
		This means that $\left(P_{i}^{C^*},P_{j}^{D^*}\right)$ should not be the optimal point for at least one line segment w.r.t. $\widetilde{\lambda_{\mathit{1}}}\left(P_{j}^{D}\right)$ or w.r.t. $\widetilde{\lambda_{\mathit{2}}}\left(P_{i}^{C}\right)$.
		For illustration, let $\tilde{l}$ denote the line segment through $\left(P_{i}^{C^*},P_{j}^{D^*}\right)$.
		
		However, it is also noticed that $\psi$ must be a closed and bounded region in one of the three cases in Fig.~\ref{feasiregion}.
		Due to the convex property of $\widetilde{\lambda_{\mathit{1}}}\left(P_{j}^{D}\right)$ or $\widetilde{\lambda_{\mathit{2}}}\left(P_{i}^{C}\right)$, there must be another point $\left(\overline{P_{i}^{C}},\overline{P_{j}^{D}}\right)$ on the same line segment $\tilde{l}$, leading to a larger value $\widetilde{\lambda_{\mathit{1}}}\left(P_{j}^{D}\right)$ or w.r.t. $\widetilde{\lambda_{\mathit{2}}}\left(P_{i}^{C}\right)$.
		That is, one of the following cases must exist:
		\begin{equation}
		\label{proofposition2}
			\widetilde{\lambda_{\mathit{1}}}\left(\overline{P_{j}^{D}}\right)>\widetilde{\lambda_{\mathit{1}}}\left(P_{j}^{D^*}\right) \quad \text{or}\quad \widetilde{\lambda_{\mathit{2}}}\left(\overline{P_{i}^{C}}\right)>\widetilde{\lambda_{\mathit{2}}}\left(P_{i}^{C^*}\right).
		\end{equation}
		Meanwhile, since $\left(P_{i}^{C^*},P_{j}^{D^*}\right)$ and $\left(\overline{P_{i}^{C}},\overline{P_{j}^{D}}\right)$ are on the same line segment, this leads to another fact that $\sigma_{i}^{C}\overline{r_{i}^{C}}$ (in the left case of \eqref{proofposition2}) or $\sigma_{j}^{D}\overline{r_{j}^{D}}$ (in the right case of \eqref{proofposition2}) gets the same value at the two points.
		Combined with \eqref{rewrittedobjective}, clearly, either of the cases contradicts the assumption that $\left(P_{i}^{C^*},P_{j}^{D^*}\right)$ is the optimal solution in $\psi$.
		This completes the proof.

	\end{appendices}

	\bibliographystyle{IEEEtran}
	\bibliography{main}
\end{document}